\documentclass[twocolumn]{aa}

\usepackage{graphicx}
\usepackage{txfonts}
\usepackage{natbib}

\bibpunct{(}{)}{;}{a}{}{,} 

\newcommand {\hi} {H\,{\small I}\,}
\newcommand {\km} {km s$^{-1}$\,}
\newcommand {\ci}{$^{\circ}$\,}
\newcommand {\mo}{$M_{\odot}$\,}

\begin{document}

\title{\ion{H}{i} study of the warped spiral galaxy NGC\,5055: 
 a disk/dark matter halo offset?}
\author{Giuseppina Battaglia\inst{1} 
\and Filippo Fraternali\inst{2} 
\and Tom Oosterloo\inst{3} 
\and Renzo Sancisi\inst{4}
       }

\institute{Kapteyn Astronomical Institute, Postbus 800, 9700 AV, 
Groningen, The Netherlands
\and Theoretical Physics, University of Oxford, 1 Keble Road, OX1 3NP,
Oxford, UK
\and ASTRON, Postbus 2, 7990 AA, Dwingeloo, The Netherlands
\and INAF-Osservatorio Astronomico, via Ranzani 1, 40127, Bologna, Italy
and Kapteyn Astronomical Institute, University of Groningen, P.O.Box 800 
The Netherlands}

\date{Received / Accepted }

\abstract{ 
We present a study of the \ion{H}{i} distribution and dynamics 
of the nearby spiral galaxy NGC\,5055 based on observations
with the Westerbork Synthesis Radio Telescope. 
The gaseous disk of NGC\,5055 extends out to
about 40 kpc, equal to 3.5 $R_{25}$ ,  
and shows a pronounced warp that starts at the 
end of the bright optical disk ($R_{25}=$ 11.6 kpc). 
This very extended warp has large-scale symmetry, 
which along with the rotation period of its outer parts 
($\simeq$ 1.5 Gyr at 40 kpc), suggests a long-lived phenomenon.
  
The rotation curve rises steeply in the
central parts up to the maximum velocity ($v\rm_{max}\simeq$ 206
\km). Beyond the bright stellar disk ($R_{25}$), it shows a
decline of about 25 \km and then remains flat out to the
last measured point. The standard analysis with luminous and dark 
matter components shows the dynamical importance of the disk. 
The best fit to the rotation curve is obtained with a ``maximum disk''.
Less satisfactory fits with lighter disks help to set  
a firm lower limit of 1.4 to the mass-to-light 
ratio in $F$ band of the disk. Such a ``minimum disk'' 
contributes about 60\% of the observed maximum rotational velocity.
  
NGC\,5055 shows remarkable overall regularity and symmetry. 
A mild lopsidedness is noticeable, however, both in the distribution 
and kinematics of the gas. The tilted ring analysis of the velocity 
field led us to adopt 
different values for the kinematical centre and for the systemic
velocity for the inner and the outer parts of the system. 
This has produced a remarkable result: the kinematical and geometrical 
asymmetries disappear, both at the same time. These results
 point at two different dynamical regimes: an inner
region dominated by the stellar disk and an outer one, dominated by
a dark matter halo offset with respect to the disk. 
         }  
\titlerunning{\ion{H}{i} study of the warped spiral galaxy NGC\,5055}

\maketitle

\section{Introduction} \label{sec:intro}
\begin{figure*}[!ht]
\begin{center}
\includegraphics[width=130mm]{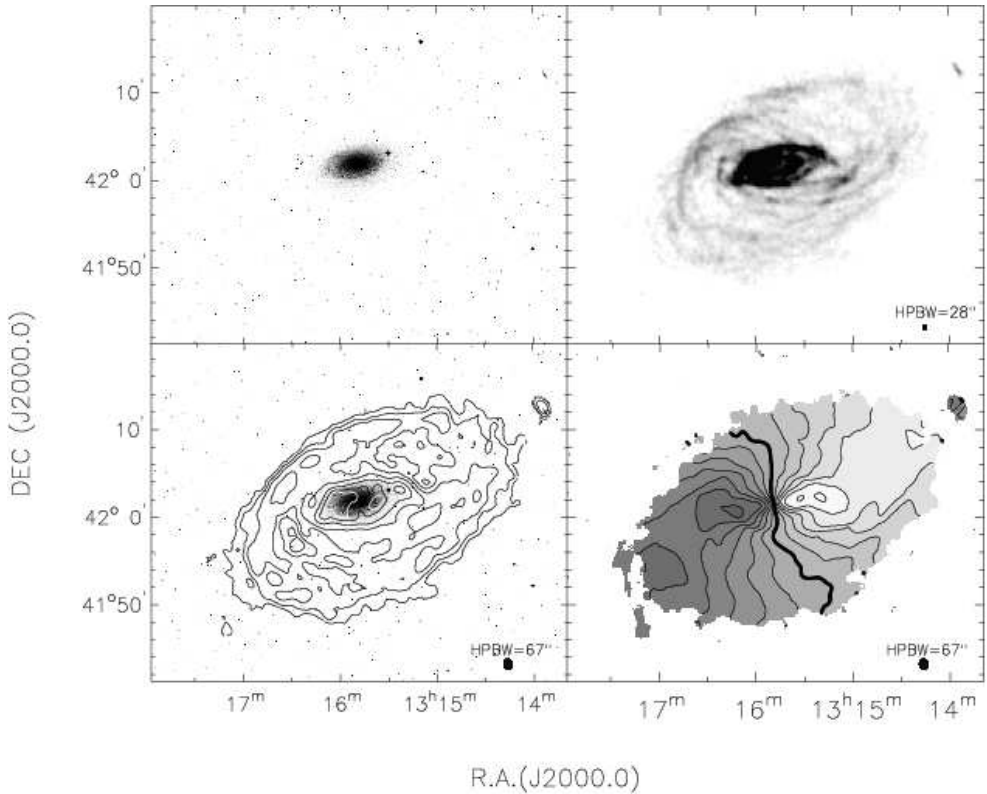}
\caption{Top left: Optical image (DSS) of NGC\,5055. Top right: 
Total \hi\ map at 28$''$ resolution. Bottom left: Total \hi\ map at 67$''$ 
resolution overlaid with the DSS optical image; contours are 3.0, 6.0, 
12.0, 20.6, 41.2, 82.4, 123.4$\times10^{19}$  cm$^{-2}$. Bottom
right: 
Velocity field at 67$''$ resolution; contours are separated by 30 
km s$^{-1}$, and the thick line shows the systemic velocity 
(497.6 km s$^{-1}$); the receding side is S-E.}
\label{fig:mosaic}
\end{center}
\end{figure*}

NGC\,5055 is an Sbc spiral galaxy (de Vaucouleurs et al. 1976) at a 
distance of 7.2 Mpc (Pierce 1994). At this distance 1$'\simeq$ 2.1 kpc. 
Figure \ref{fig:mosaic} (top left) shows the DSS image. The main 
optical and radio parameters are listed in Table \ref{tab:data5055}.

Previous 21-cm line observations (Bosma 1978) have shown the presence of an
extended, warped gaseous disk and a declining rotation curve.
In this study we used new Westerbork Synthesis Radio Telescope (WSRT) 
observations to study 
the warp and the dynamics of the system.

The warp starts at the end of the bright stellar disk, around $R_{25}$, 
and is exceptionally extended and symmetric. 
This suggests a stable dynamics that precludes recent formation. Warps
are still poorly understood phenomena. They are very common and ubiquitous 
 (e.g. Garcia-Ruiz, Sancisi \& Kuijken 2002 and references therein), 
but it is not clear yet whether this means that they are 
easily excited or that there is a mechanism capable of maintaining them. 
Several different possibilities have been proposed to explain 
their origin 
such as: tidal interactions (Burke 1957; Kerr 1957; Hunter \& Toomre 1969), 
misalignment between disk and halo (Debattista \& Sellwood 1999), 
 discrete bending modes due to a tilted oblate dark matter halo 
(Sparke \& Casertano 1988), 
effects of the intergalactic medium and accretion (Kahn \& Woltjer 1959; 
Jiang \& Binney 1999). In particular, the possible role of dark 
matter halos has been widely considered. 
However, none of these explanations has been completely accepted.

The rotation curve of NGC\,5055 shows a decline at the end of the bright 
optical
disk. Both rotation curves with a similar shape and correlations with the 
luminosity and light distribution have been known to exist  
(Casertano \& van Gorkom 1991; Persic and Salucci 1991; Broeils 1992). 
They seem to point at rules in the distribution and relative dynamical 
contributions of the luminous and dark components. In particular, velocity 
declines may be indications of a dominant contribution of 
the stellar disk in the inner parts and of a transition to a dark halo in 
the outer parts.

NGC\,5055 also shows the presence of gas rotating with velocities lower than 
the rotational velocities of the disk. Such "anomalous" gas has already been 
detected in several other spiral galaxies seen at various 
inclination angles (e.g. Fraternali 
et al. 2002; Boomsma et al. 2005) and has been interpreted as gas 
residing in their halo regions (Swaters, Sancisi \& van der Hulst 1997).   
 
The new high sensitivity and high resolution \hi\ observations of NGC\,5055
with the WSRT allow us to investigate all the aspects mentioned above. 

\section{Observations and data analysis} \label{sec:obs}

We observed NGC\,5055 with the DZB spectrometer of the Westerbork
Synthesis Radio Telescope. The WSRT observing parameters are given 
in Table \ref{tab:observations}.
Calibration and data reduction were done using standard procedures of the 
MIRIAD (Multichannel Image Reconstruction, Image Analysis and Display) 
package (Sault et al. 1995).    
Twenty-two channels free of line emission were identified at the low
velocity end of the band and twenty-four at the high velocity end. 
The radio continuum emission was obtained by interpolating with a straight 
line and then subtracted from the line channels. Three data cubes at
different angular resolutions were obtained. The full resolution 
(18$''\simeq$ 627 pc) data cube was obtained using all baselines. 
For the data cubes at resolutions of 
28$''$ and 67$''$, we used baselines shorter than 5k$\lambda$ and 3k$\lambda$, 
respectively. Table \ref{tab:cubes} lists the parameters for the
data cubes. 
A Hanning smoothing was then applied in velocity leading to a 
resolution of 16 \km. The dirty channel maps were cleaned using a
Clark CLEAN algorithm (Clark 1980).  
The final data cubes 
(at 18$''$, 28$''$, 67$''$ 
resolution) have r.m.s. noises per channel of 0.3, 0.4, 0.6 mJy/beam 
respectively. In these cubes the respective minimum detectable column
densities in one velocity resolution 
element (at 5$\sigma$) are 8.4$\times 10^{19}$  cm$^{-2}$ 
(0.67 \mo/pc$^2$), 
4.6$\times 10^{19}$  cm$^{-2}$ (0.37 \mo/pc$^2$), and 
1.2$\times 10^{19}$  cm$^{-2}$ (0.10 \mo/pc$^2$).
 
\begin{figure}[!ht]
\begin{center}
\includegraphics[width=70mm]{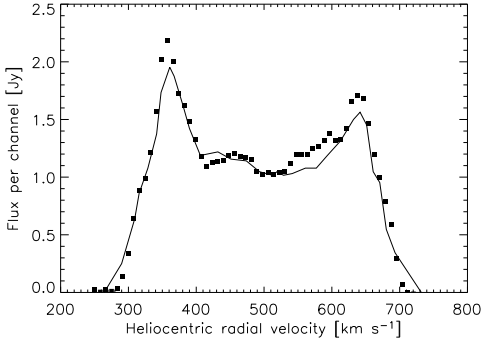}
\caption{Comparison between the global \hi\ profile of NGC\,5055 
from this work 
(squares) and 
from single dish observations (full line) by Rots (1980).}
\label{fig:profile}
\end{center}
\end{figure}

Comparison between the global \hi\ profile from these observations
 and from single 
dish measurements (Rots 1980) shows that there is no loss
 of \hi\ flux due to the missing short spacings (Fig. \ref{fig:profile}).

\begin{figure*}[ht]
\begin{center}
\includegraphics[width=130mm]{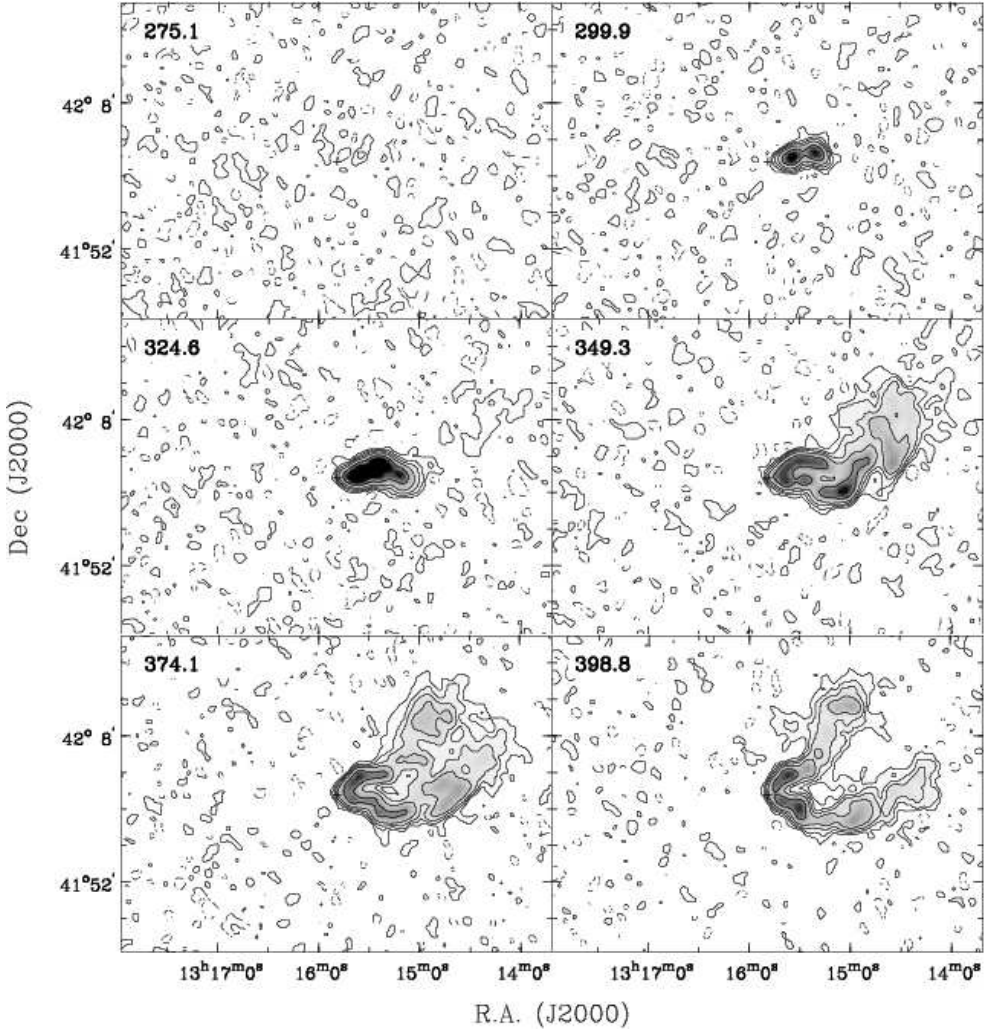}
\caption{\hi\ channel maps of NGC\,5055 at 67$''$ resolution 
(Hanning smoothed). 
Contours are $-$1.2 (dashed), 1.2, 3, 6, 12, 30, 60, 120, 300 mJy/beam; the r.m.s. 
noise is 0.6 mJy/beam. The heliocentric velocities (in \km) are shown in the 
upper left corner; the cross indicates the kinematical centre of the galaxy.}
\label{fig:channels60}
\end{center}
\end{figure*}

The cleaned data cubes were analyzed using the Groningen Image 
Processing System (GIPSY) package (van der Hulst et al. 1992). 
Channel maps at 
67$''$ resolution, with a separation of
27 \km and without primary beam correction, are shown in 
Fig. \ref{fig:channels60}. 
These channel maps show the well-known pattern of a differentially
rotating disk and a very extended warp in the outer parts.  

\subsection{Total \hi\ maps and velocity fields}

For each resolution (18$''$, 28$''$, 67$''$), total \hi\ maps were obtained by 
adding the channel maps containing neutral hydrogen emission (from
$\simeq$ 
280 \km to 
$\simeq$ 730 km s$^{-1}$). 
The area 
containing line emission was defined by visual inspection. 
The total \hi\ mass, corrected for the primary beam attenuation, is 
6.2$\pm 0.3 \times 10^9$ M$_{\odot}$. The value obtained by Bosma
(1978) is 6.3$\times 10^9$ M$_{\odot}$, 
after correction for the distance adopted in this work. For 
comparison, the single-dish mass (Rots 1980) is 5.9$\times 10^9$ M$_{\odot}$.

Figure \ref{fig:mosaic} 
shows the total \hi\ map at 28$''$ resolution (top right) and at 
67$''$ resolution overlaid with the DSS optical image (bottom left). 
The \hi\ disk extends 
out to about 3.5 $R_{25}$ (see Table \ref{tab:data5055}), equal to 40 kpc. 
The extent derived from previous observations of lower sensitivity (Bosma 1978) 
was 37 kpc. 
The high surface brightness part of the \hi\ disk (dark gray in
Fig.~\ref{fig:mosaic}) has 
the same extent (about 10 kpc) and orientation 
as the bright optical disk. The outer
parts show a 
change in orientation of about 20$^{\circ}$ as compared to the inner region due 
to the warping of the \hi\ disk (see Sect.~\ref{sec:warp}). 

Remarkably, the
extended, warped HI layer shows spiral arm features and a large-scale pattern 
(arms or ring) in the density distribution.
In the velocity field there are wiggles possibly 
associated with these spiral arms that suggest
the presence of streaming motions.

\begin{figure*}[ht]
\begin{center}
\includegraphics[width=130mm]{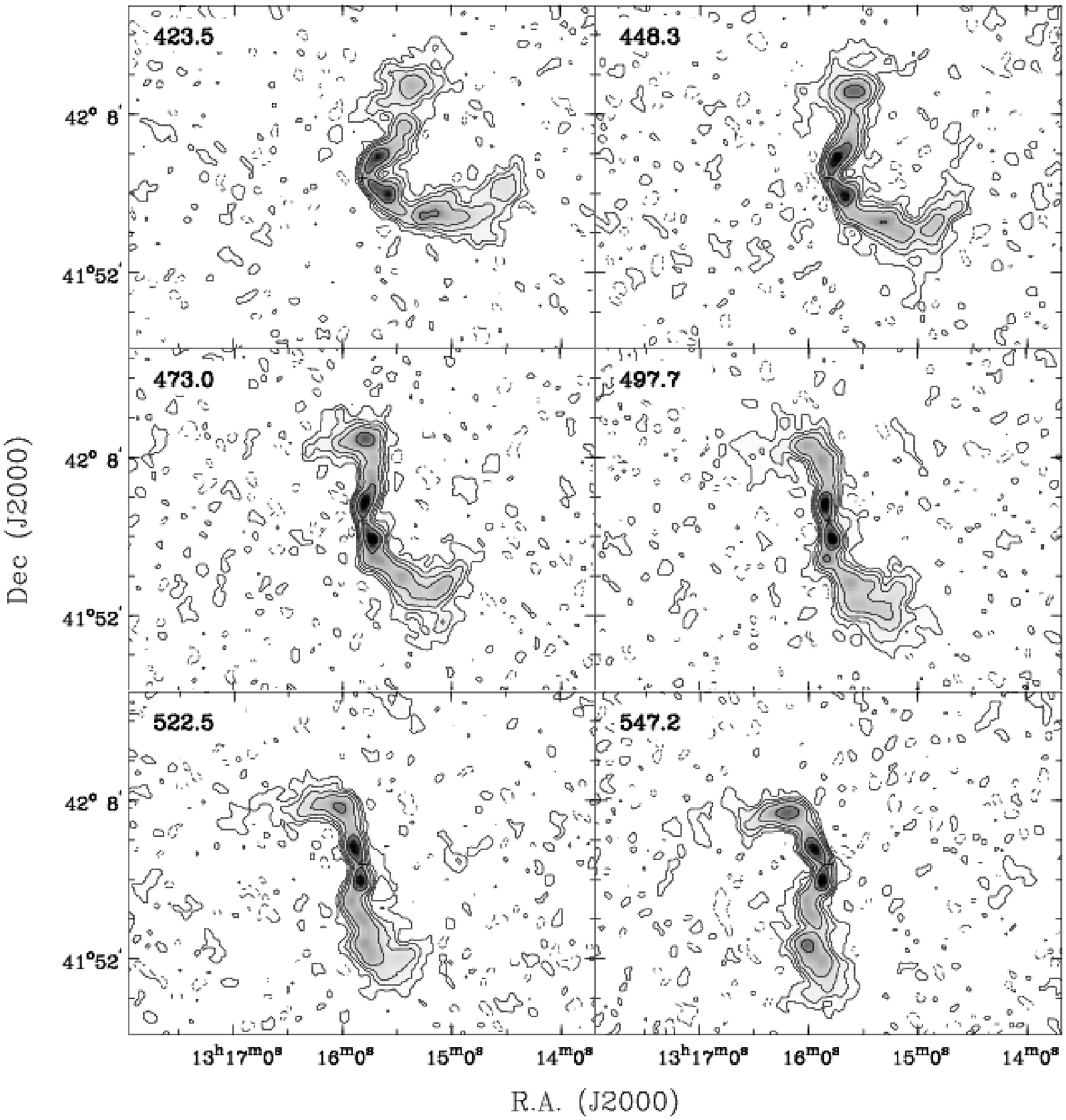}
\end{center}
Figure \ref{fig:channels60}: continued.
\end{figure*}

\begin{figure*}[ht]
\begin{center}
\includegraphics[width=130mm]{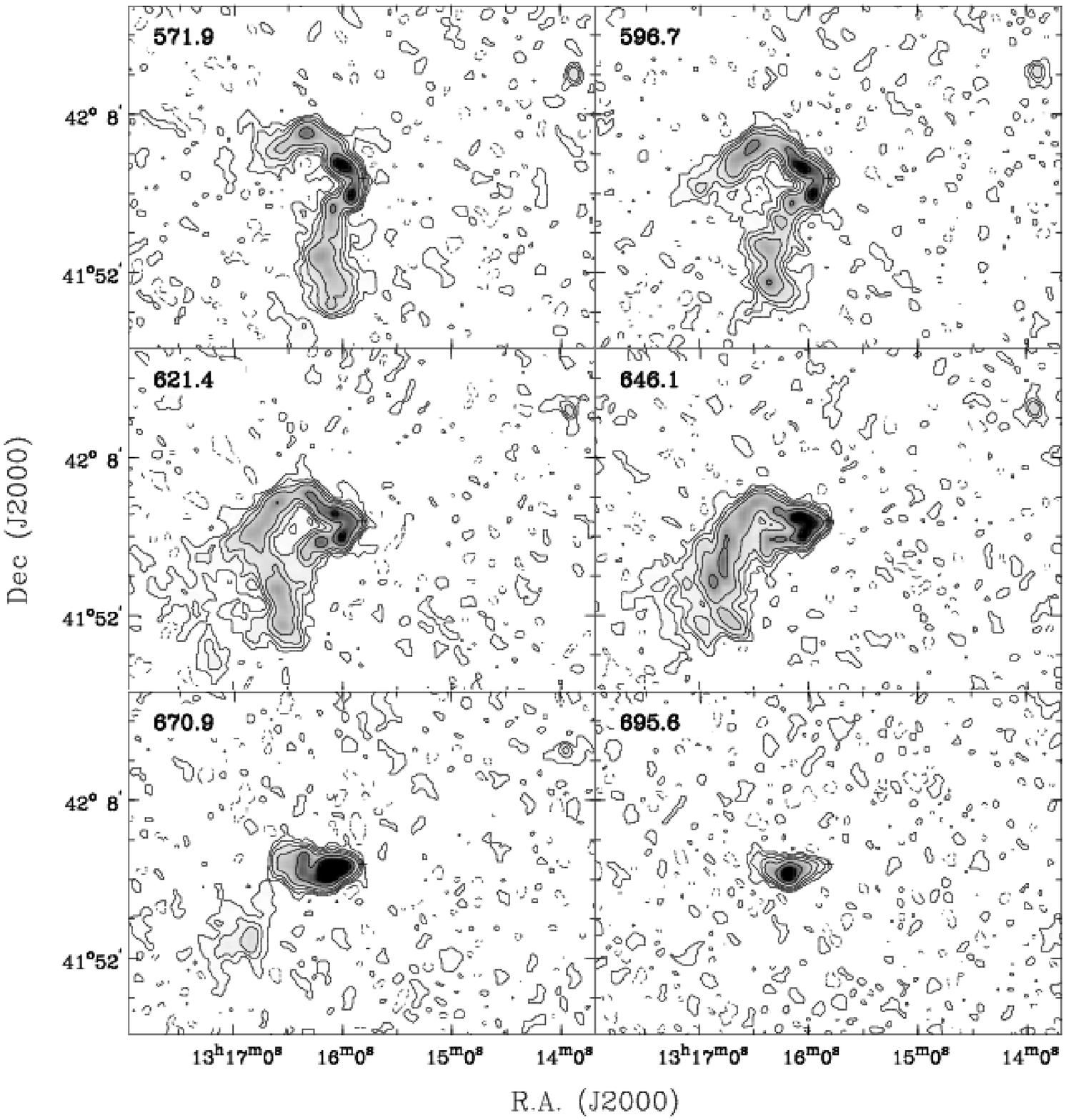}
\end{center}
Figure \ref{fig:channels60}: continued.
\end{figure*}

Velocity fields for each of the three resolutions were obtained by fitting 
Gaussians
to the velocity profiles. This method gives good results except in the 
very central region ($R\leq$ 18$''=$ 627 pc),  
where the velocity profiles are affected by beam smearing.  
Figure~\ref{fig:mosaic}  (bottom right) shows the radial velocity field at 
low resolution (67$''$). 
This velocity field is very symmetric and, in the outer parts, shows the 
characteristic 
behaviour of a kinematical warp: the position angle of the kinematical
major axis varies 
with the radius, from a value of about 100$^{\circ}$ within $R\rm_{Ho}$ to 
about 120 $^{\circ}$ in the outer region. Wiggles are visible
especially in the S-W side.

\subsection{UGC 8313}

We also detected neutral hydrogen emission from a companion galaxy of
NGC\,5055, UGC\,8313, a small galaxy of type SBc (Tully 1988) at 
$\alpha=$ 13$^h$13$^m$54.1$^s$, $\delta=$ 42$^{\circ}$12$'$35.7$''$ 
(2MASS, 2000), at a projected distance of 50 kpc to the North-West of NGC\,5055 
(Fig. \ref{fig:mosaic} and
channel maps from 571.9 \km to 670.9 \km, Fig. \ref{fig:channels60}). 
By integrating the \hi\ channel
maps at 18$''$ resolution from $\simeq$ 555 \km to $\simeq$ 687 \km, we
obtained a total \hi\ mass, corrected for the primary beam attenuation,
of 7$\times$10$^7$ M$_{\odot}$. In comparison, Bosma \cite{bosma} found a total
\hi\ mass of 2.8 $\times$10$^7$ M$_{\odot}$ (with a distance of 7.2
Mpc). 
From the global \hi\ profile,
we obtained a systemic velocity, $v\rm_{sys}\simeq$ 620 \km. The spatial resolution 
of our data was too low and the object too small ($R\simeq$1$'\simeq$2.1 kpc) 
to derive a detailed rotation curve. But taking the 
effects of bandwidth and the
uncertain inclination angle (the galaxy seems almost edge-on) into account, we set 
a firm lower limit for the rotational velocity, $v\rm_{max}\ge$ 50 \km. 
This value was derived both from the global
\hi\ profile and from the position-velocity diagram along the major
axis. 
We estimated the total mass of UGC\,8313 to be $M=1.2\times10^9$\mo, which
represents a lower limit because we do not observe 
the flat part of the rotation curve 
and we do not know the exact extent of the disk.

\section{Determination of the rotation curve} \label{sec:rotcurve}

In order to obtain the kinematical parameters and the rotation curve for NGC\,5055, 
a tilted ring model (Begeman 1987) was fitted to the observed velocity fields 
at different resolutions. The tilted ring model assumes that a spiral galaxy can be 
described by a set of 
concentric rings. Each ring is characterized by two orientation angles 
(position angle $\phi$ and 
inclination angle $i$) and the circular velocity $v\rm_C$. Every ring
is assumed to have 
the same value for the dynamical centre and the systemic velocity. At 
each resolution, the rings were fitted with a radial increment 
of the corresponding beam-width. The points were weighted by the cosine of the 
azimuthal angle with respect to the major axis. 

An iterative procedure was followed. To determine the dynamical 
centre, only the points within 11.6 kpc (corresponding to $R_{25}$) in the velocity 
field at full resolution were used. While keeping this position of the centre 
fixed, the systemic velocity was fitted. Figure \ref{fig:tilted} shows that the
systemic velocity is not constant. The average value within 230$''$ ($\simeq$ 8 kpc) 
is $\simeq$ 492~\km, while in the outer parts 
it is $\simeq$ 500~\km. This may suggest a dynamical decoupling 
of the central part with respect to 
the outer part of the galaxy (see Sect.~\ref{sec:symmetry}). 
The mean systemic velocity (line
in Fig. 
\ref{fig:tilted}) was estimated as the weighted mean 
of the points at full resolution within the bright optical disk (11.6
kpc). 
Points within 60$''$ 
($\simeq$ 2 kpc) were 
excluded in the fitting 
procedure because of beam-smearing and insufficient statistics. 
The values found for the kinematical centre and for the 
systemic velocity are reported in Table \ref{tab:data5055}.

We held the above parameters (centre and systemic velocity) fixed and 
we fitted first the position angle and 
then the inclination angle. The  
full-resolution data were used for the region within $\simeq$ 400$''$
(14 kpc), and
for larger distances we used 
the data at lower resolution. The position
angle has a constant 
value 
($\simeq$ 100$^{\circ}$) in the inner part; between $R_{25}$ (11.6
kpc) and $R\rm_{Ho}$ (16.7 kpc) it 
decreases by 10$^{\circ}$ and further out rises 
to 120$^{\circ}$ at the last measured point (Fig. \ref{fig:tilted},
top right). The trend found by Bosma \cite{bosma}
is very similar: he gives an average value of 99$^{\circ}$ for the inner part 
and 115$^{\circ}$ for the outer part. The inclination angle shows an 
almost constant value of 64$^{\circ}$ out to $R_{25}$. Beyond that region it 
decreases to 50$^{\circ}$ and then increases in the outer parts (mean value 
$\simeq$ 55$^{\circ}$). Because of the large errors in the fit of the inner part, 
the value of the inclination angle was fixed to 63$^{\circ}$ within
30$''$ (1 kpc). 
Bosma assumed a constant value of 55$^{\circ}$ for the whole galaxy; 
Pierce at al. (1994), with photometric CCD observations, found 
64$^{\circ}$ for the inner part.    

\begin{figure*}
\begin{center}
\includegraphics[width=130mm]{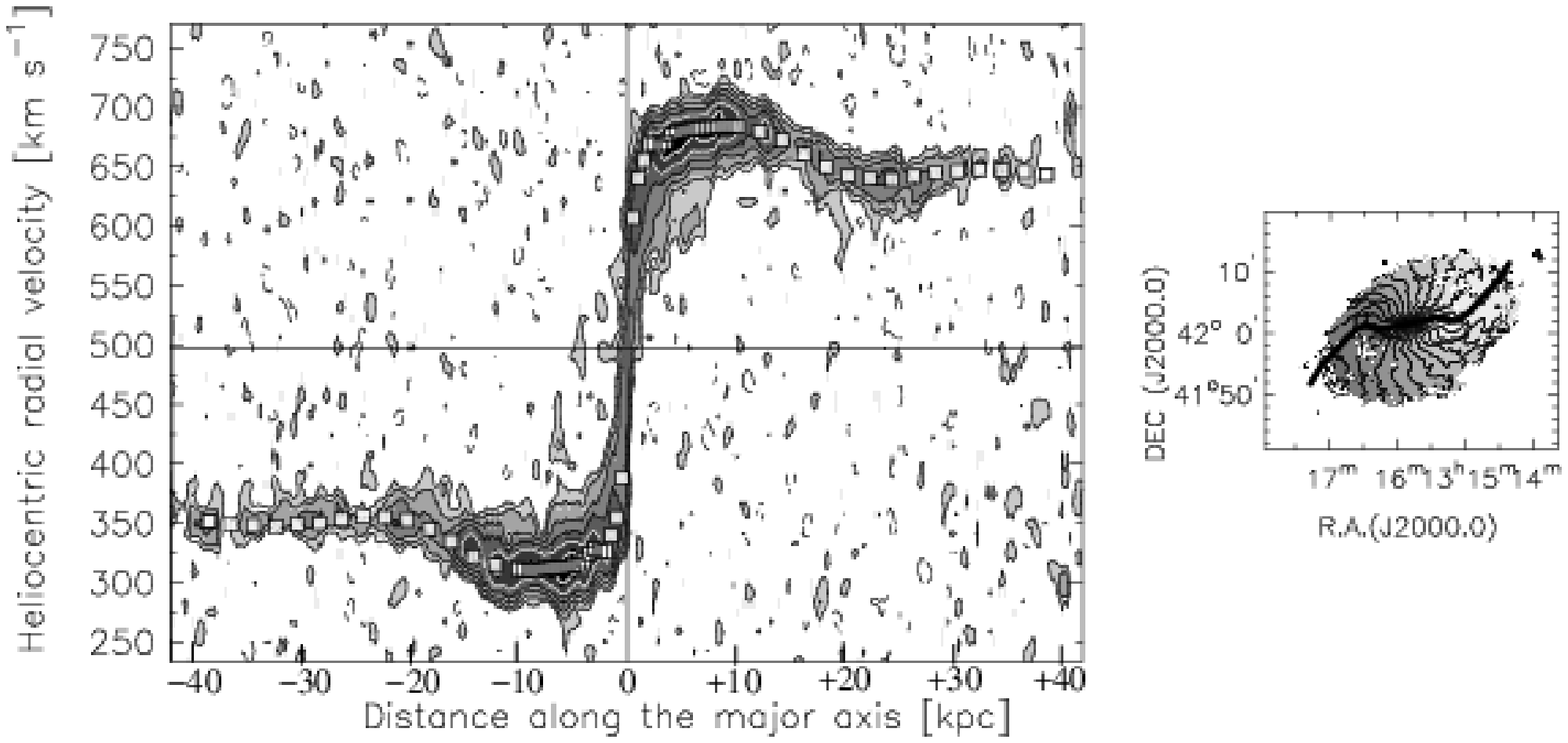}
\caption{Left: Position-velocity diagram along the path (major axis) 
shown on the velocity
field at 28$''$ resolution on the right (black thick line). 
The spatial resolution is 28$''$, and the velocity resolution is $\sim$ 
16 km s$^{-1}$. Contours are -2, 2, 4, 8, 16, 32, 64, 128, 256 $\sigma$, where 
$\sigma=$ 0.4 mJy/beam. The horizontal line shows the systemic
velocity. The white squares show the projected circular velocity.}
\label{fig:curve30_xv}
\end{center}
\end{figure*}

\begin{figure*}[ht]
\begin{center}
\includegraphics[width=60mm]{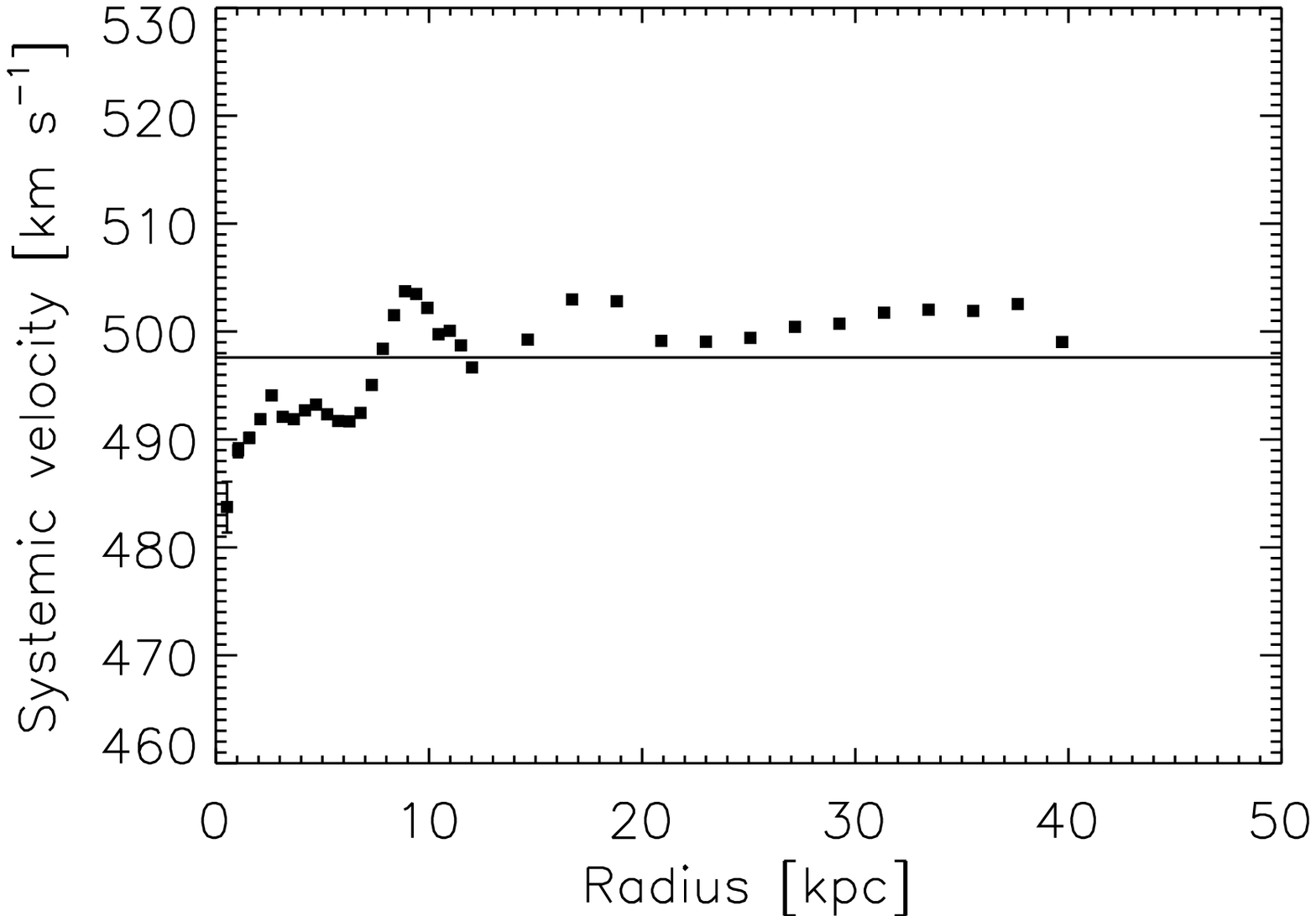}
\includegraphics[width=60mm]{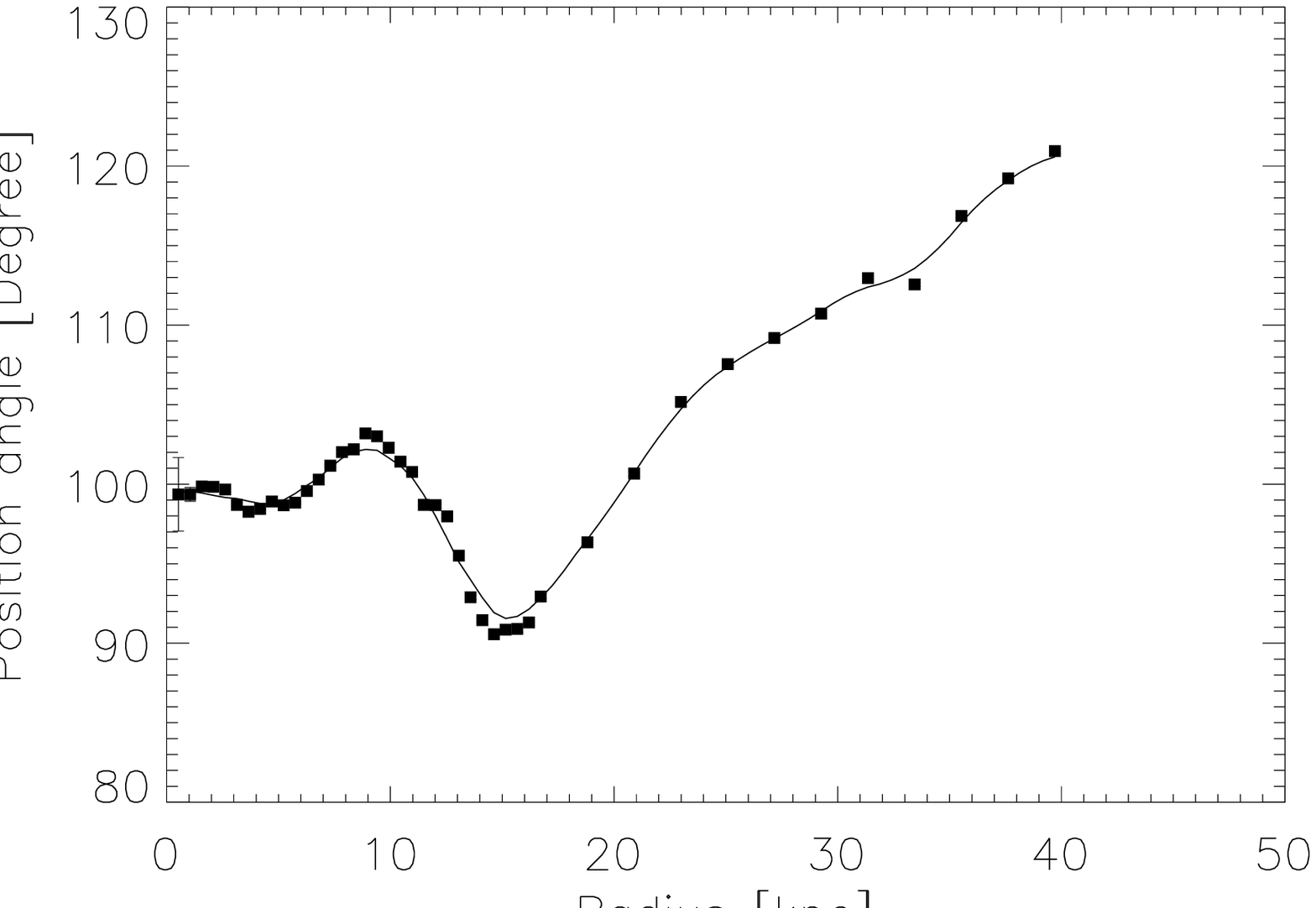}
\includegraphics[width=60mm]{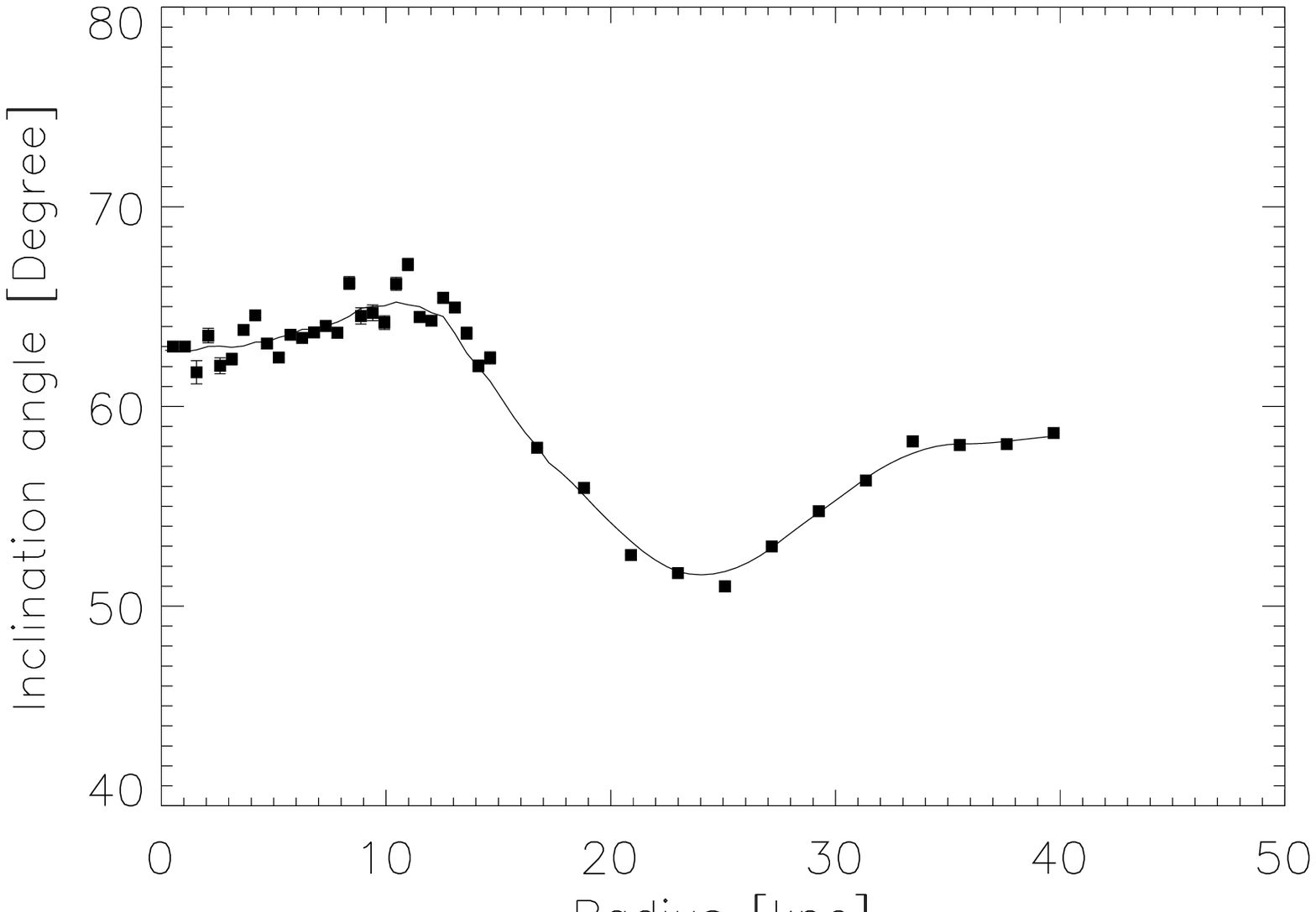}
\includegraphics[width=60mm]{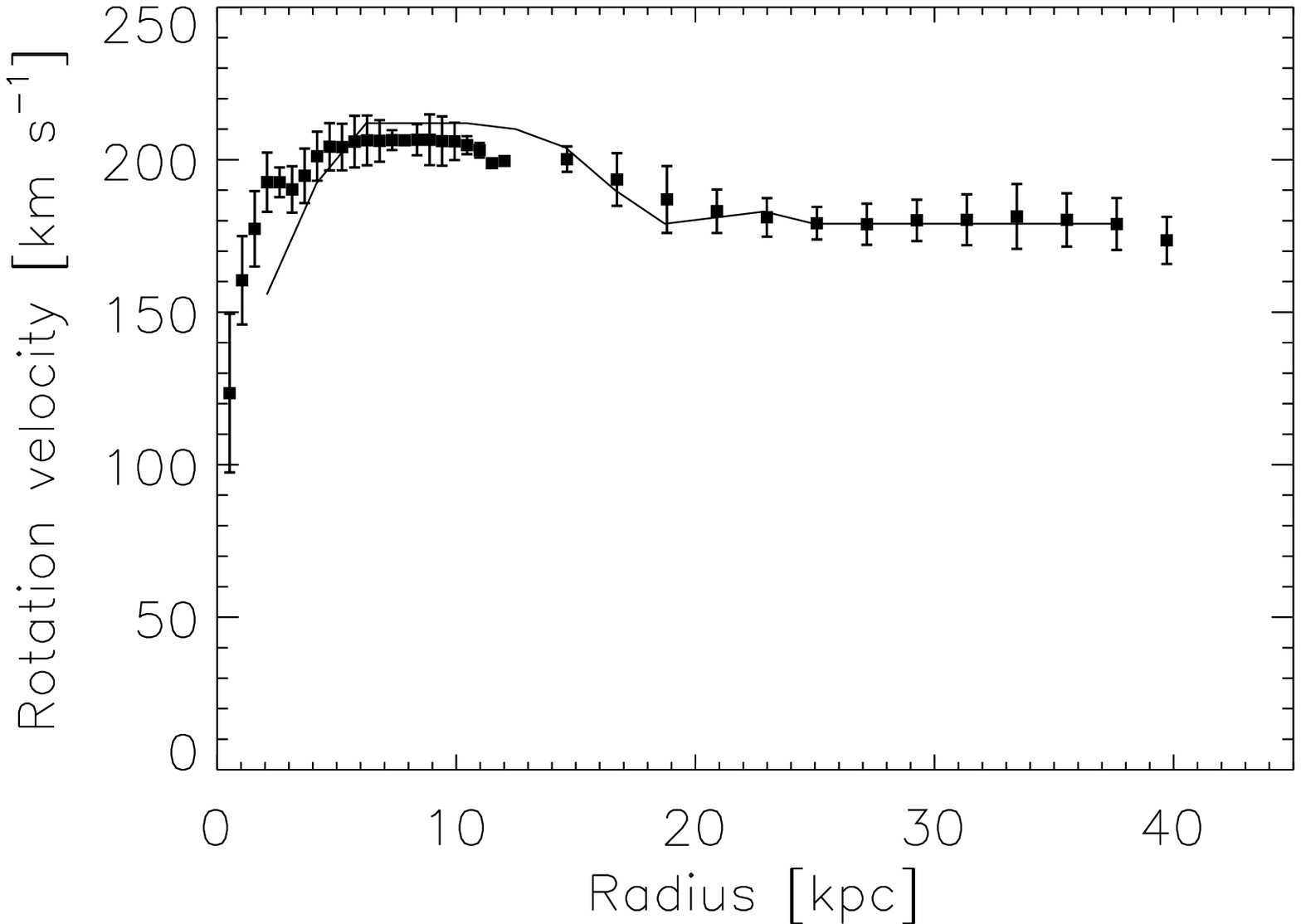}
\caption{Parameters for NGC\,5055 obtained from the tilted ring model fit. 
In the top left panel the horizontal line shows 
the mean systemic velocity adopted in this work, $v\rm_{sys}$=497.6 km s$^{-1}$. 
The full lines in the plots for the position and inclination angles were obtained 
smoothing the data, in order to remove small-scale fluctuations.  
Bottom right panel: rotation curve from 
this work (filled squares) compared with the rotation curve (line) 
derived by Bosma (1978).} 
\label{fig:tilted}
\end{center}
\end{figure*}

Finally, after having determined all the above parameters, we derived
the circular velocity. 
The rotation curve (Fig. \ref{fig:tilted}, bottom right) reaches 
the maximum velocity, $v\rm_{max}=$ 206 km s$^{-1}$ between  
1.9$h$ and 2.7$h$ ($h=$ 3.4 kpc).  
Between 10 and 20 kpc, it decreases by about 25 km s$^{-1}$
 and then becomes almost constant out to the last measured point 
($R_{\rm out}=$19$'\simeq$ 40 kpc). The mass of NGC\,5055 within 
40 kpc derived from the circular velocity at that radius is 
$M(<R_{\rm out})\simeq 2.7 \times 10^{11}$ \mo.
The error bars show the differences
in circular velocities between approaching and receding side and
include, therefore, the effect of deviations from circular motions.  
The solid line shows the rotation curve derived by Bosma \cite{bosma}. In the 
inner regions (inside 5 kpc), the circular 
velocities obtained in this work are larger than the rotation velocities 
derived by Bosma (maximum difference 40 km s$^{-1}$). This difference can be 
explained with the lower resolution of Bosma's data ($\simeq$ 60$''$) 
and the consequent beam-smearing effects. 
Bosma obtained a 
maximum velocity of 214 km s$^{-1}$; this larger value is explained by 
the different inclination used to de-project the observed radial velocity 
($i\rm_{Bosma}=$ 55$^{\circ}$, $i\rm_{our}=$ 63$^{\circ}$). The quality of
the derived rotation curve can be tested by overlaying it on the
data. Figure \ref{fig:curve30_xv} (left) shows a position-velocity diagram
taken at the positions of the highest projected velocities (actual
major axis) as shown in the right panel of Fig. \ref{fig:curve30_xv}. From
this p-v diagram it is clear that the derived rotation curve follows
the peak of the line profiles over the whole galaxy closely.      

\section{Warp} \label{sec:warp}

The channel maps and the velocity field of NGC\,5055 show the
characteristic pattern of a kinematical warp. In 
both the \hi\ map and the velocity field (Fig.~\ref{fig:mosaic}), the 
orientation of the major axis changes with distance from
the centre, varying 
from 100$^{\circ}$ within $R_{25}$ to 120$^{\circ}$ in the outer region.

\begin{figure}[!h]
\begin{center}
\includegraphics[width=70mm]{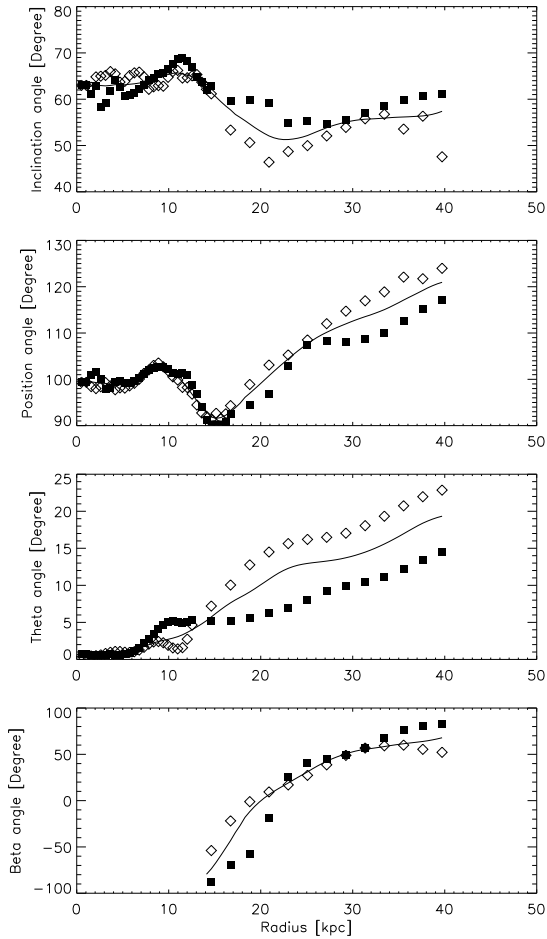}
\caption{Radial profiles of the geometrical parameters of the warp 
of NGC\,5055. 
Filled squares show the approaching side, open diamonds the receding
side, and the 
full 
line shows the parameters 
for the whole galaxy.}
\label{fig:parwarp}
\end{center}
\end{figure}

In order to investigate the geometry of the disk and the symmetry of the 
warp, the two orientation angles (position and inclination angle) were 
determined for the 
approaching and the receding sides separately (Fig.~\ref{fig:parwarp}).  
The position angles of the two halves of the 
galaxy (Fig.~\ref{fig:parwarp}, second panel) are similar: 
the only difference is in the outer region, where the receding side 
has values that are 15$^{\circ}$ higher than the approaching side. The behaviour 
of the inclination angle is similar for the two sides 
(Fig.~\ref{fig:parwarp}, first panel) within 400$''$ 
($\simeq$ 14 kpc). Between 14 kpc and 28 kpc there are large differences, since 
on the approaching side, the inclination angle is almost constant around 
57$^{\circ}$, whilst on the receding part it decreases from 60$^{\circ}$ 
to 47 $^{\circ}$.     

These orientation angles are set with respect to the plane of the
``sky''. In order to display the space orientation of the rings, 
we used a reference frame defined by the plane of the inner 
galaxy (within 
$R_{25}$). We determined the angle between the plane of every ring and the 
plane of the inner disk, $\theta$, as well as the angle between the line of 
nodes and the line of intersection of the inner disk with the plane of
the ``sky'', 
$\beta$ (Schwarz 1985). 
The trend of the $\theta$ angle shows that the inclination of the
rings is 
almost 
constant within the stellar disk ($R\lesssim$ 10 kpc), and then it increases
monotonically with 
distance from the centre; the outermost ring has an inclination of 
20$^{\circ}$ with respect to the central plane. 
There is an evident systematic difference between the two sides. 
The behaviour of the $\beta$ angle suggests 
that the line of nodes changes direction monotonically with radius.
 
\begin{figure}[!h]
\begin{center}
\includegraphics[width=70mm]{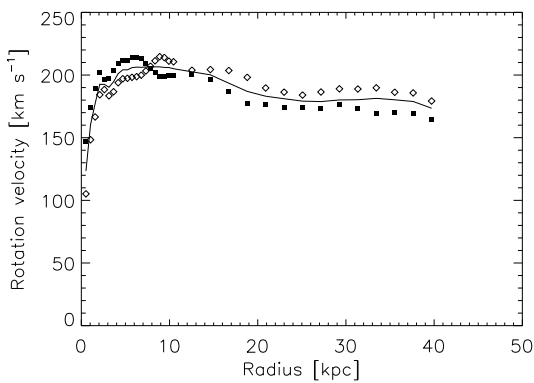}
\caption{Comparison between the rotation curves for NGC\,5055 derived
for the approaching
side (filled 
squares) 
and for the receding side (open diamonds). The full line shows the 
rotation curve for the whole galaxy.}
\label{fig:curva_apprec}
\end{center}
\end{figure}

\begin{figure}[!h]
\begin{center}
\includegraphics[width=60mm]{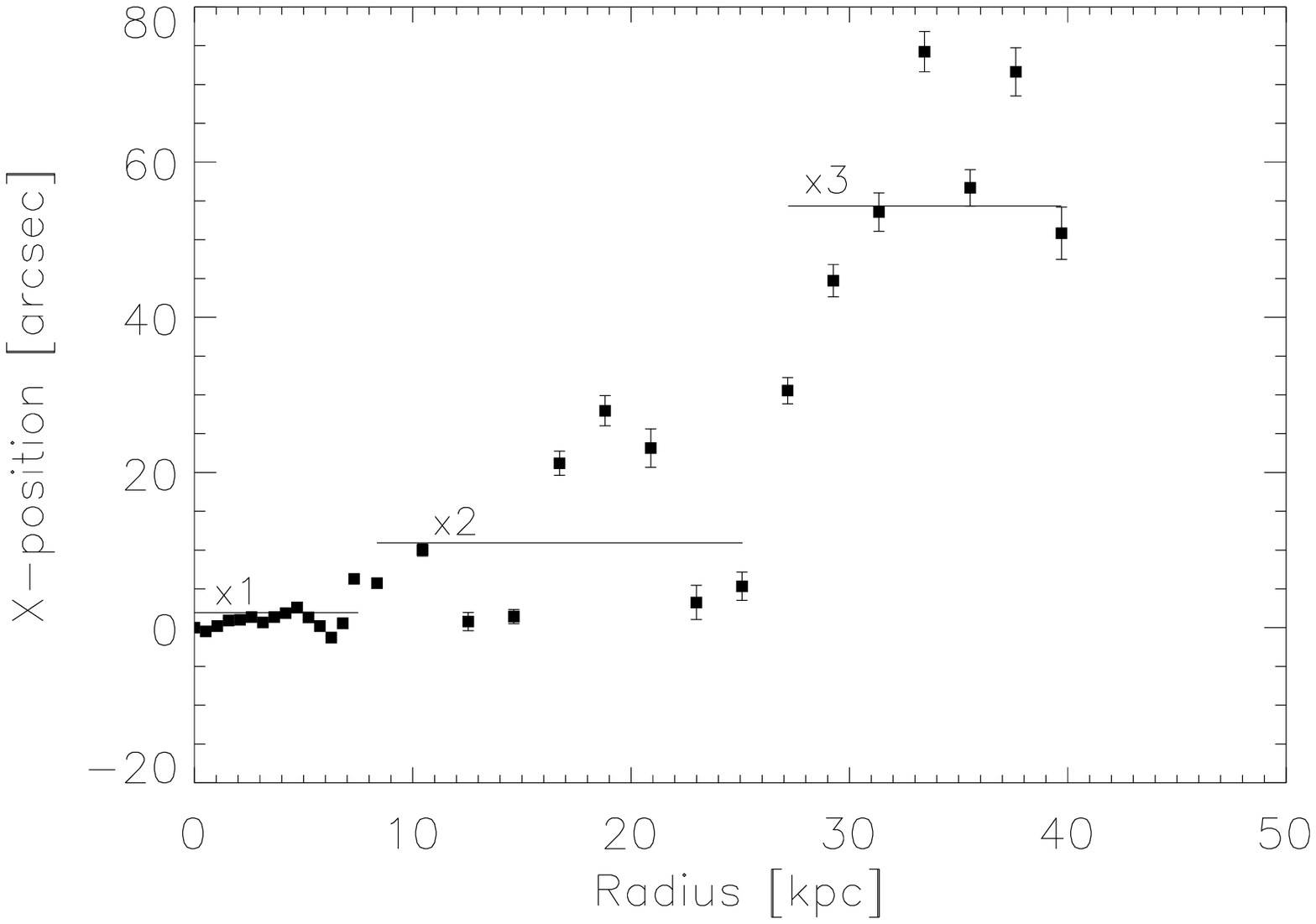}
\includegraphics[width=60mm]{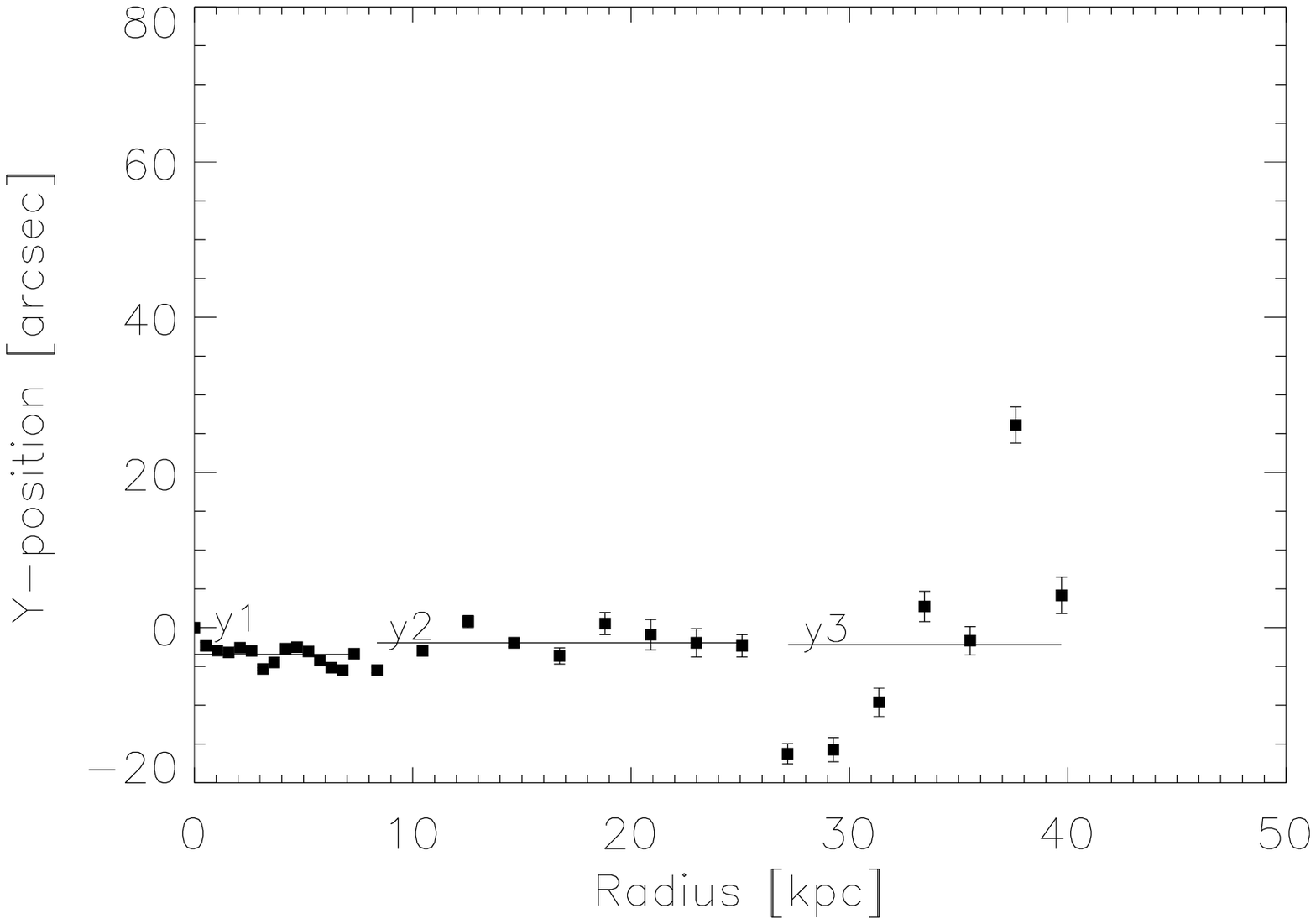}
\caption{Radial trend of the $x$ (top) and $y$ (bottom) position of
the dynamical centre 
for NGC\,5055. The horizontal lines indicate the three different values for the 
coordinates used in the tilted ring model fit described in Sect.~\ref{sec:symmetry}.}
\label{fig:centre}
\end{center}
\end{figure}

\section{Symmetry of the rotation curve} \label{sec:symmetry}

In Sect.~\ref{sec:rotcurve} we derived the rotation curve for the whole galaxy (Fig. 5).
Here we investigate the symmetry of the curve by analyzing the approaching and
the receding side separately. For this we have used the geometrical parameters 
for the approaching and the receding side as derived in the previous section. 
Figure \ref{fig:curva_apprec} shows that the two halves of the galaxy have 
similar trends of the circular velocity, and the two curves have the
same overall shape as the 
rotation curve derived for the whole galaxy. They show, however, 
a systematic difference with
the rotational velocity higher on the approaching side within 
$R\sim$ 8 kpc and 
lower beyond that radius. The differences are typically 
15 to 20 \mbox{\km}. The radius $R=$ 8 kpc
seems to represent a turnover point for the dynamical behaviour of
the two halves of the galaxy. As described in Sect.~\ref{sec:rotcurve}, 
we also noticed that at about the same distance
from the centre 
a sudden change of the systemic velocity occurs (see
Fig. \ref{fig:tilted}, top left); 
the value in the inner part is about 492 km s$^{-1}$, but beyond $R=$ 8 kpc it 
becomes $v\rm_{sys}\sim$ 500 km s$^{-1}$. This suggests that the
kinematics of NGC\,5055 could be better described by taking two 
different values for the systemic velocity: a lower value in the inner part 
and a higher one in the outer region. This may also imply different
values for the centre of the inner and outer rings.    

In the classic approach to deriving of a rotation curve with a
tilted ring fit, the centre and the systemic velocity of the galaxy
are kept fixed while deriving the other parameters (inclination angle,
position angle, and rotation velocity). 
In view of the peculiar behaviour of the systemic velocity 
(Fig.~\ref{fig:tilted}, 
upper panel) and of the rotation curves of the approaching and
receding sides (Fig.~\ref{fig:curva_apprec}), we attempted to
relax this procedure and to let the position of the centre and the systemic
velocity vary. We first fitted the observed velocity
field, allowing the kinematical centre to be free and 
fixing the systemic velocity 
at 492 km s$^{-1}$ within 8 kpc ($\sim 230''$ radius) and at 500 km
s$^{-1}$ beyond that. We found that the {\it x} position of the centre 
changes considerably from the inner to the outer rings, whilst
the $y$ coordinate does not show any significant variation. 
For simplicity, we took three constant values for the {\it x} position. 
Figure \ref{fig:centre} shows the values of the coordinates we adopted and
the respective intervals. Then we repeated the fitting procedure
keeping these two values for the systemic velocity and the three 
values for the centre as fixed. 

Figures \ref{fig:parwarp_gr} and \ref{fig:curva_apprec_gr} show 
the geometrical  parameters and the rotation curves obtained with this
procedure. The 
symmetry between the two halves of NGC\,5055 is now striking, and 
the differences visible in 
Fig. \ref{fig:parwarp} and \ref{fig:curva_apprec} have disappeared. 
The rotation curve for the whole galaxy 
does not show any significant difference from the rotation curve 
obtained with fixed kinematical centre and systemic velocity, 
but the rotation curves for the approaching and 
receding sides now overlap almost completely. A discussion of these
results is given in Sect.~\ref{sec:symm_disc}.
Figure \ref{fig:ellissi} shows the tilted ring model for NGC
5055 obtained with the parameters of the new fit.

It is interesting to note that similar variations of the systemic velocity
have also been found in NGC\,628 (Kamphuis \& Briggs 1992) and in M101 
(Kamphuis 1993).

\section{Mass models}

The rotation curve of NGC\,5055 (Fig. 10) was derived with the tilted ring 
model fit 
described in Sect.\,\ref{sec:rotcurve}. 
Here it is used for the study of the mass distribution in 
NGC\,5055. We consider a mass model with three components - the stellar disk, 
the gaseous disk, and the dark matter halo - and derive the velocity 
contributions 
that best represent the observed rotation curve through a least-square fit
 using a GIPSY routine.
 For estimating the stellar disk 
contribution, we used two different sets of photometric data in the ``F'' band.
 
For radii less than 7.2$'$ ($\simeq$ 15 kpc), we took the CCD brightness
profile published by Kent \cite{kent}, and  
photographic data out to 10.3$'$ ($\simeq$ 21.5 kpc)
were available from Wevers \cite{wevers}. In order to match the two data sets, 
we added 0.2 
magnitudes/arcsec$^2$ to Wevers' data. Such a
correction is normal when comparing photographic and CCD data (Kent 1987a). 
To take the presence of the warp into consideration, we deprojected the 
brightness 
profile by using the inclination derived in Sect.\,\ref{sec:rotcurve}. 
Figure \ref{fig:brightness} (top) shows the deprojected brightness profile.

The radial \hi\ profile, shown in Fig. \ref{fig:HIdensity}, was 
derived from the observed surface density (total \hi\ map) 
using the parameters from the tilted ring fit. The profile shows a 
density drop-off in the region between 10 and 15 kpc (around 
$R_{25}$) at the location where the warp starts. 
For estimating the contribution of the gaseous disk to 
the rotation curve, the \hi\ surface density was multiplied by a factor of 1.4, 
to take the helium abundance into account. We adopted a scale height of 0.2 
kpc for the gaseous disk and 0.4 kpc for the stellar disk. 
The velocity contribution of the stellar component was 
calculated from the formula published by Casertano \cite{casertano83}. 
The available 
optical photometry extends out to 10.3$'$ ($\simeq$ 21.5 kpc), while the last 
measured point for our \hi\ data is at 19$'$ ($\simeq$ 40 kpc). 
The contribution of the stellar component was calculated in the 
hypothesis of both an exponential disk and a drop in the 
brightness profile beyond 10.3$'$. No significant differences were
found between the two cases. As usually done, the $M/L$ ratio for 
the stellar component was assumed to be independent of radius. 

Figure \ref{fig:brightness} (bottom) shows the contribution of the
stellar component to the observed rotation velocity obtained 
by maximizing the stellar disk contribution in order 
to match the observed rotation curve in the inner parts (maximum disk 
hypothesis).
It is clear from the large discrepancy between
observed and predicted curves that a dark matter component is needed 
starting from well inside $R_{25}$ to account 
for the observed rotational velocities.

As mass models for the dark matter halo, we adopted isothermal and  
NFW density profiles, as described in the following subsections.

\subsection{Dark matter halo: isothermal profile}

First we modeled the dark matter halo with a quasi-isothermal density profile: 
$$\rho(r)=\rho_{0}[1+({r\over{r_C}})^2]^{-1}$$
where $\rho_{0}$ is the central density and $ r_C$ is the core radius. 
The circular velocity due to this mass distribution is
$$V{\rm_c}^2(r) = V{\rm_c}^2(\infty) (1 -  \frac{r_c}{r} {\rm arctg}\frac{r}{r_c})$$
where $V{\rm_c}^2(\infty)= 4 \pi G \rho_0 r\rm_c^2$ is the asymptotic rotational velocity. 
The parameters that best fit the data are $\rho_0= 7.6\pm 1.9 \times 10^{-3}$ \mo pc$^{-3}$ , 
$r\rm_c= 8.4 \pm 1.4 $ kpc, and 
$M/L= 3.8\pm 0.1$ in $F$ band, corresponding to a $M/L$ 
in $B$ band of about 3.2\footnote{We assumed $M_{\odot,B}=$ 5.48 and $M_{\odot,F}=$ 4.5}. 
The above corresponds to a maximum disk fit, and gives a mass for the 
disk component within the last measured point $M_*(<R_{\rm out})= 8\times 
10^{10}$ \mo, and for the dark matter halo $M_{\rm DM}(<R_{\rm out})= 1.9\times 
10^{11}$ \mo.

This fit (Fig. \ref{fig:decomp}, top left) 
reproduces the observed circular velocities in detail out to 
the region of the maximum velocity and, in particular, the feature at  
2 kpc radius. The decline 
at the end of the bright optical disk is also reproduced. In the outer 
region, there is a good agreement between observed and predicted curves.

The top right panel in Fig. \ref{fig:decomp} shows the fit in the minimum 
disk hypothesis (the contribution of the dark matter halo is maximized), 
where $\rho_0= 6.2\pm 1.0$ \mo pc$^{-3}$  and  
$r\rm_c= 0.30 \pm 0.03 $ kpc. 
For an acceptable fit, it is not possible to decrease
the contribution of the disk further. This sets a firm lower limit to the
stellar mass-to-light ratio: $M/L\gtrsim$ 1.4. With such a value for the $M/L$ 
ratio, the luminous matter is responsible for about 60\% of the observed 
circular velocity at the peak of the rotation curve. For such a fit, 
$M_*(<R_{\rm out})= 3 \times 10^{10}$ \mo and 
$M_{\rm DM}(<R_{\rm out})= 2.5\times 10^{11}$ \mo.

\subsection{Dark matter halo: NFW density profile}

For the dark matter halo, we also used the universal density profile 
derived by Navarro, Frenk, \& White (1995, 1996)
$$\rho(r) = \frac{\delta_c \rho_c^0}{c \frac{r}{r_v}(1+ c \frac{r}{r_v})^2}$$ 
with circular velocity
$$V_{\rm c}^2(r) = {{V\rm{_v}}^2 g(c) r_v \over{r}}\biggl[\,{\rm ln}(1+c\,\frac{r}{r_v})-{c\,\frac{r}{r_v}\over{1+c\,\frac{r}{r_v}}}\biggr]$$
where $\delta\rm_c$ is the density contrast at the virial radius, 
$c$ the concentration, $r\rm_v$ the virial radius, and $v\rm_v$ the 
circular velocity at the virial radius. We use the concentration and the 
virial radius as free parameters for the dark matter halo. 
 Figure
\ref{fig:decomp} 
(bottom left) shows 
the fit in the maximum disk hypothesis ($M/L=$3.6, $r\rm_v= 316\pm 23$ kpc, $c=6.3\pm0.8$). 
This gives a virial mass of  $M_{\rm vir}= 1.5\times 10^{12}$ \mo and a disk and 
dark matter mass within the last measured point,  
 which are very similar to the combination of a maximum disk and 
an isothermal halo.
The value of the concentration, however, appears very 
low compared to the one predicted by CDM simulations
(e.g. Wechsler et al. 2002).  
If we keep the $M/L$ free in the fit, we obtain $M/L=$ 3.5, 
very close to the maximum disk hypothesis.

When trying to minimize the disk contribution in the fit, 
the values of the concentration tend to increase considerably. 
Since the typical values for a DM halo of galactic mass range from 10 to 20, 
we set an upper limit of 20 to the concentration. In this way a minimum 
disk fit corresponds to a maximum concentration fit. 
Figure \ref{fig:decomp} (bottom right) shows the result of the 
maximum concentration fit ($c=$20, $M/L=2.8\pm 0.1$, $r\rm_v= 219\pm 4$ kpc).   
The quality of the fit is not 
as good as in the maximum disk. In particular, the feature at 2 kpc radius 
is not reproduced and in the outer region a discrepancy with the
observed velocities begins to show up. 
For a further discussion see Sect.~\ref{sec:disc}.  

\section{Halo gas}

Recent \hi\ observations of edge-on spiral galaxies (Swaters, Sancisi 
\& van der Hulst 1997; Matthews \& Wood 2003)
have revealed the presence of extra-planar gas located at several kpc 
above and below the galaxy plane and with a rotational velocity that is lower than the gas in
the disk. Such peculiar kinematics made it possible to also detect such a halo
\hi\ in less inclined systems (e.g. NGC\,2403, Fraternali et al. 2002), 
where it shows up in p-v diagrams along the major axis as a wing at low
rotational velocities.

We investigated the presence of such gas in NGC\,5055 
by analyzing both the channel maps and the
position-velocity diagrams. 
Figure \ref{fig:curve30_xv} (left) shows a p-v
diagram at a resolution of 28$''$ along the major axis. In the inner
parts of this diagram, the line profiles are not
symmetric but have ``wings'' extending towards the systemic
velocity (low rotational velocity). 
In the receding side the ``lagging'' gas is mostly visible
within about 7 kpc from the centre and shows a maximum projected 
difference of about 100 \km from the rotational velocity. In the
approaching side, there is a hint of very high velocity gas up to
$\sim$ 200 \km of a projected difference from rotation. 
In both cases the emission comes from regions
inside the bright optical disk. Our observations, with an integration time of 24 hours, 
were not deep enough
to allow a complete analysis of the halo gas. So we did not perform any separation
of this gas from the ``cold'' thin disk.      

\section{Discussion} \label{sec:disc} 

The distribution and dynamics of the neutral gas in the nearby spiral
galaxy 
NGC\,5055 were studied using
new high-sensitivity WSRT observations.
In this section we discuss the main results.

\subsection{\hi\ density distribution and warp} \label{sec:warp_disc}

The \hi\ disk of NGC\,5055 extends out to 19$'$ (40 kpc) from the
centre, far beyond the optical radius ($R_{25}$=5$'$.5).
The outer parts of the \hi\ disk are strongly warped as indicated
by the total \hi map and the velocity field in Fig.~\ref{fig:mosaic}.
In the tilted ring model for the disk, the outer rings are warped by 
about 20\ci with respect to the
inner ones ($\theta$ angle in Fig.~\ref{fig:parwarp_gr}, panel 3 from the top).
The warp begins at about 5$'$ (10 kpc) at the end of the bright optical disk.
This behaviour is commonly observed in spiral galaxies (Briggs 1990;
Garcia-Ruiz et al. 2002). This is also close to the edge of
the bright \hi\ disk where the \hi\ surface density drops from 5.5 to about 1 
\mo pc$^{-2}$ (Fig.~\ref{fig:HIdensity}).
Further out, the density is below 1 \mo pc$^{-2}$  and continues to decrease 
slowly by about 0.06 \mo pc$^{-2}$kpc$^{-1}$. 
The density distribution of \hi\ in the region of the warp is not homogeneous, and
spiral-like features are clearly visible in the high resolution 
total \hi\ map (Fig.~\ref{fig:mosaic}, top-right panel).

Despite the mild asymmetries between the approaching and the receding
sides described in Sect.~\ref{sec:symmetry}, 
the warp of NGC\,5055 is highly symmetric both in geometry (see
Figs.~\ref{fig:parwarp} and \ref{fig:parwarp_gr}) and 
in kinematics, as is clearly shown by the velocity field (Fig.~\ref{fig:mosaic}, 
lower right panel).
Such a symmetry is remarkable if one considers the exceptional extent
of the warp itself.
The orbital time (roughly a lower limit to the life time of the warp) at
$R=$ 40 kpc is about 1.5 Gyrs.
This suggests that the warp of NGC\,5055 is a long-lived phenomenon.

Galactic warps have been known to exist for several decades, but a
satisfactory dynamical explanation has not been found yet.
The dark matter halo may play an important role, as suggested by the 
empirical rule that warps start at the end of
the stellar disks where the dark matter potential becomes
dynamically dominant (Briggs 1990).
Several mechanisms have been proposed for the formation (and/or
maintenance) of warps, such as accretion of intergalactic material with
different angular momentum (Jiang \& Binney 1999), interactions with
companion galaxies (Hunter \& Toomre 1969), or misalignment between the
disk and the dark halo (Debattista \& Sellwood 1999). 

In NGC\,5055, the warp tends to align the position angle of the outer
disk towards the companion galaxy UGC\,8313 (see Fig. 1). 
This may suggest a role of the latter in the formation and/or maintenance of the warp.
However, the motion of UGC\,8313 is retrograde with respect to that of the disk. 
This implies that the dynamical effect of UGC\,8313 is likely to be small. Moreover, 
given the large difference in mass, one can expect that NGC\,5055 would have a greater 
influence on UGC\,8313, which appears undisturbed. Therefore, the interaction 
between the two galaxies cannot be strong.

\subsection{The dynamics} \label{sec:dyn_disc}

The rotation curve of NGC\,5055 (Fig.~\ref{fig:curva_apprec_gr}) 
rises steeply in the
inner parts, reaching the maximum (206 \km) at 4 kpc (2$'$) from
the centre. 
Between 10 and 20 kpc, it shows a decline of about 25 \km, and beyond 20
kpc, it remains almost flat out to the last measured point (at 40
kpc).
Two features of this rotation curve deserve particular attention: 
one is the inner bump at about 2 kpc from the centre and the
other is the decline at 10$-$20 kpc. 

\begin{figure}[!ht]
\begin{center}
\includegraphics[width=70mm]{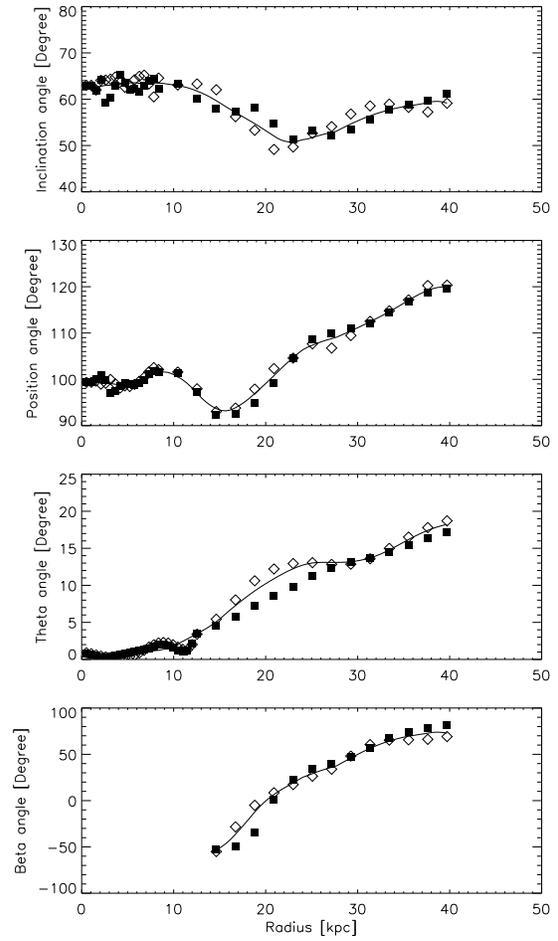}
\caption{Radial profiles of the warp parameters for NGC\,5055, obtained using three values for 
the kinematical centre and two values for the systemic velocity (see Sect.~\ref{sec:symmetry}). 
Filled squares 
show the approaching side, open diamonds the receding side, and the full line shows the parameters 
for the whole galaxy. Note the remarkable symmetry of the galaxy if
compared with the same plots in Fig.~\ref{fig:parwarp}.}
\label{fig:parwarp_gr}
\end{center}
\end{figure}

\begin{figure}[!ht]
\begin{center}
\includegraphics[width=70mm]{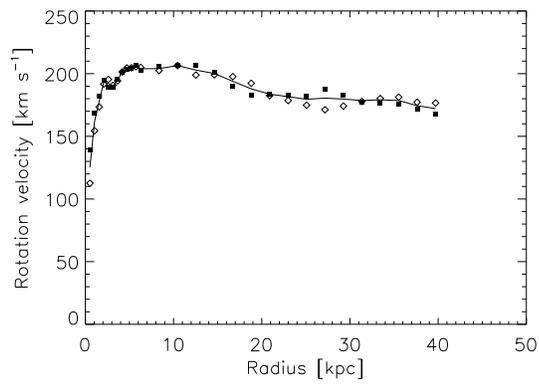}
\caption{Comparison between the rotation curve for the approaching side (filled squares) 
and receding side (open diamonds). The full line shows the rotation 
velocities obtained for the whole galaxy. 
We obtained these rotation curves using three values for 
the kinematical centre and two values for the systemic velocity (see Sect.~\ref{sec:symmetry}). 
The two rotation curves are remarkably symmetric
(compare with Fig.~\ref{fig:curva_apprec}).}
\label{fig:curva_apprec_gr}
\end{center}
\end{figure}

\begin{figure}[!ht]
\begin{center}
\includegraphics[width=70mm]{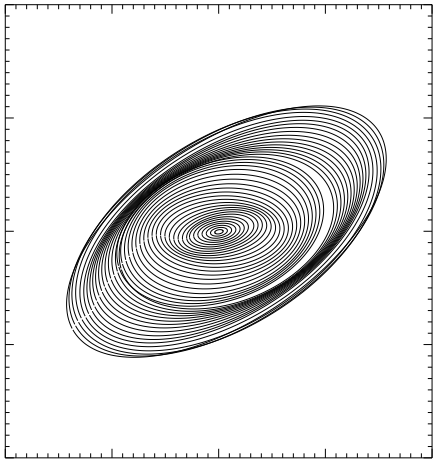}
\caption{Tilted ring model for NGC\,5055 obtained from the parameters
shown in Figs. \ref{fig:centre} and \ref{fig:parwarp_gr}.}
\label{fig:ellissi}
\end{center}
\end{figure}

\begin{figure}[!ht]
\begin{center}
\includegraphics[width=70mm]{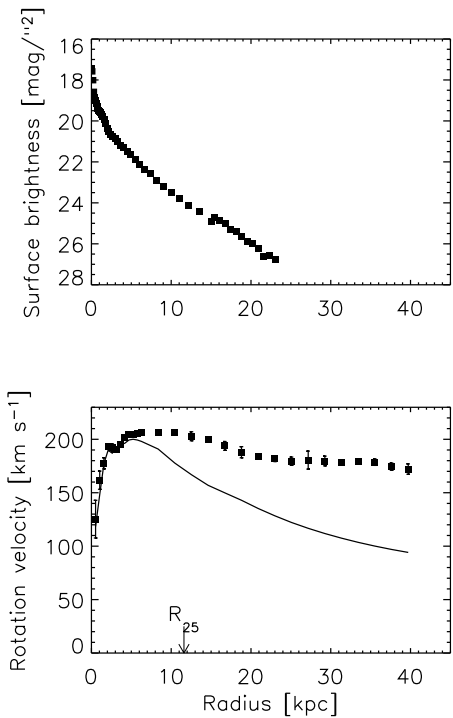}
\caption{Top: optical surface brightness profile ({\it F} band) for NGC\,5055. Data 
within 15 kpc from the centre are from CCD observations (Kent 1987); 
the data in the outer parts are from photographic plates (Wevers 1984). 
Bottom panel:observed \hi rotation curve (filled squares with error bars) and 
rotation curve calculated from the optical surface 
brightness profile (line).}
\label{fig:brightness}
\end{center}
\end{figure}

The inner bump is observed in both the
approaching and the receding sides (Fig.~\ref{fig:curva_apprec_gr}). 
The excess velocity in the bump is of the order of 10 \km.
There is no doubt that the bump is real (cf. Fig. 4), because 
it occurs in a region where both inclination and position
angles are almost constant (Figs.~\ref{fig:tilted},
\ref{fig:parwarp}, \ref{fig:parwarp_gr}) 
and is present in both halves of the
disk (Figs.~\ref{fig:curva_apprec} and \ref{fig:curva_apprec_gr}). 
Similar features are often observed in the inner regions of spiral
galaxies (Sancisi 2004).
A comparison with the photometric profile of NGC\,5055 (Fig.~\ref{fig:brightness}) shows
that there is a corresponding bump in surface
brightness within 2 kpc of the centre. 
Indeed, the maximum disk fit to the rotation curve reproduces 
the inner shape of the rotation curve (Fig.~\ref{fig:decomp}, left
panels) rather
accurately, whereas fits with dominant smooth halo components do not (Fig.~\ref{fig:decomp}, 
right panels). 
This strongly suggests that, in the inner part of the galaxy, the
dominant component is the stellar one or is a dark component
distributed like the stars.
In other words, the stars trace the gravitational potential 
in the inner parts.  

As to the decline beyond 10 kpc, is it real or could it be  
the result  of an imperfect model fitting
of the warped disk? 
It is clear that correct determination of the two parameters (inclination
and position angles) that describes the projected geometry of the warp
is crucial for deriving correct values for the rotation velocity. 
Let us first consider the position angle.
The kinematical major axis of a rotating disk is the location where 
the projected rotation velocities reach their highest values. 
When the position angle varies with {\it R}, the major axis
is not a straight line but is ``S'' shaped if the warp is
symmetric. 
In Fig.~\ref{fig:curve30_xv} (right), 
the shape of the actual major axis of NGC\,5055 is
plotted over a velocity field at 28$''$ resolution. 
In the left panel of Fig.~\ref{fig:curve30_xv}, 
the p-v plot along the ``S'' shaped
major axis is shown with the projected rotation curve of the galaxy overlaid (white squares). 
The squares follow the peaks of the line profiles very closely, and the
small discrepancies are mainly due to asymmetries between the
approaching and the receding sides. 
Thus the fitted position angles for each ring do represent the location
of the maxima in the velocity field. 

Clearly, the decline in
the rotation curve is not caused by an error in the determination of
the position angle. 
Now consider the inclination angle.
An error in determiningof the inclination angles of the outer rings
would affect the derived rotation velocities. 
The decline of the outer rotation curve would indeed disappear if
the values of the inclination angles were wrong by about 10\ci (more
face-on) beyond 10 kpc. 
Such a high systematic error is unlikely considering the small
errors in the fit and the very small differences between the approaching and
receding sides (Figs.~\ref{fig:parwarp}, \ref{fig:parwarp_gr}). 
Moreover, a fit of the outer contours (at
$R\simeq$18$'$) of the total \hi\ map gives a value for the outer 
inclination of 57.6\ci, very close to the one obtained at the same radius 
with the tilted ring fit (Figs.~ 6, 9).
In conclusion, the declining rotation curve must be a real feature of
the dynamics of NGC\,5055. 

\begin{figure}[!ht]
\begin{center}
\includegraphics[width=70mm]{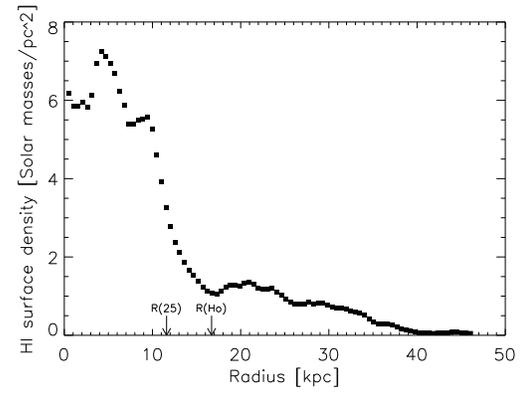}
\caption{Radial \hi profile for NGC\,5055. The arrows show the De 
Vaucouleurs ($R_{25}$) and Holmberg 
radii (R$_{Ho}$).}
\label{fig:HIdensity}
\end{center}
\end{figure}

\begin{figure*}[ht]
\begin{center}
\includegraphics[width=60mm]{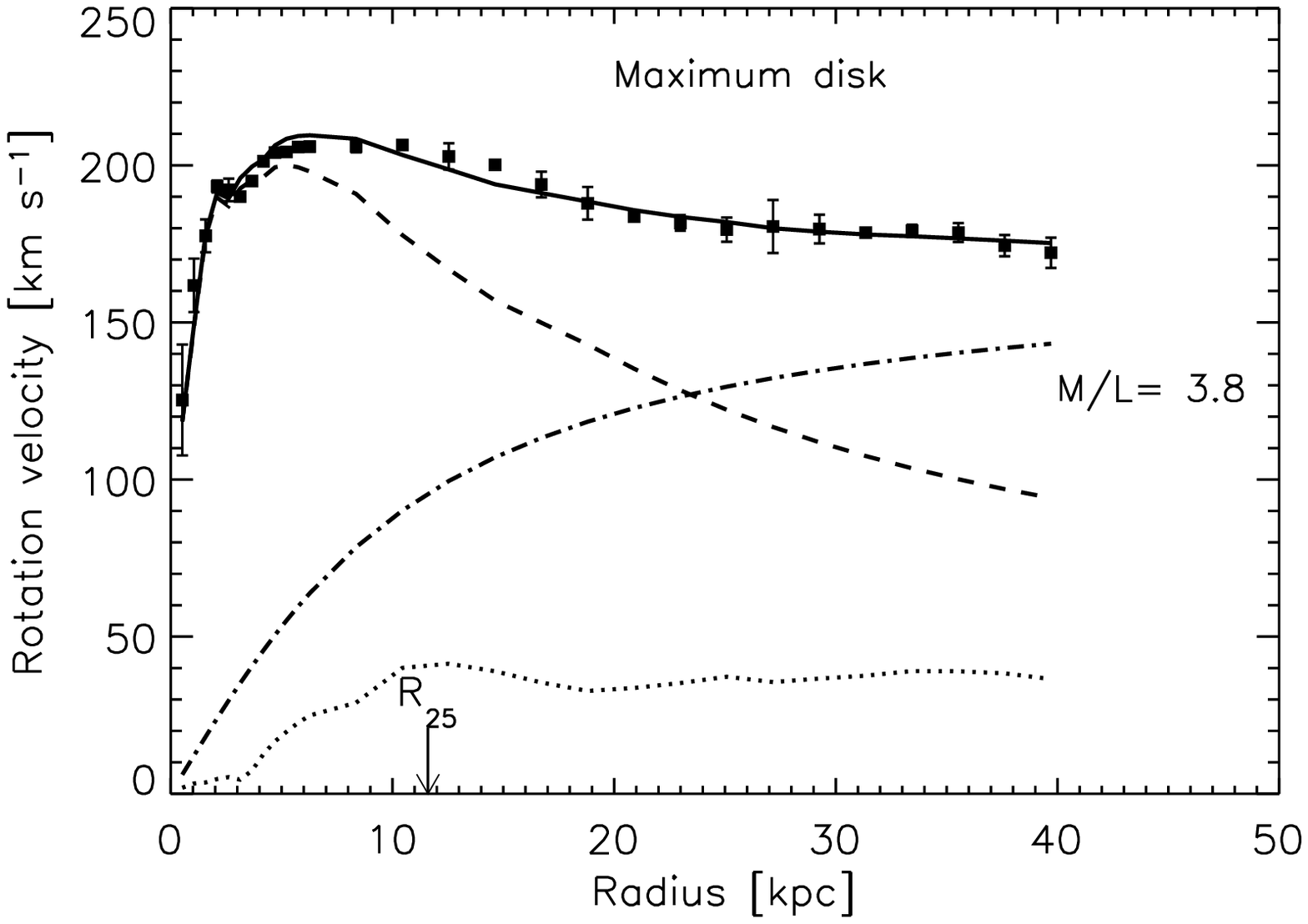}
\includegraphics[width=60mm]{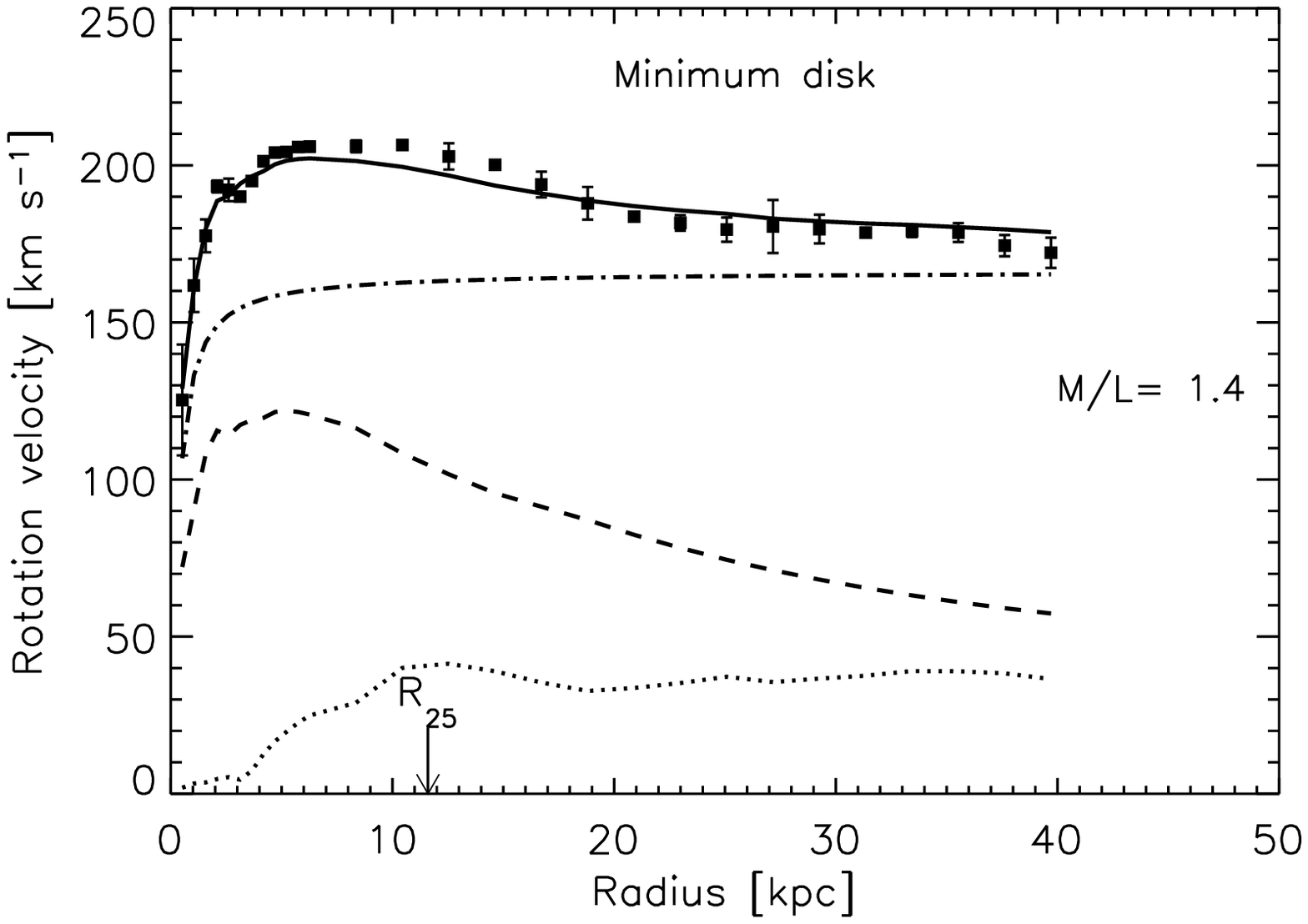}
\includegraphics[width=60mm]{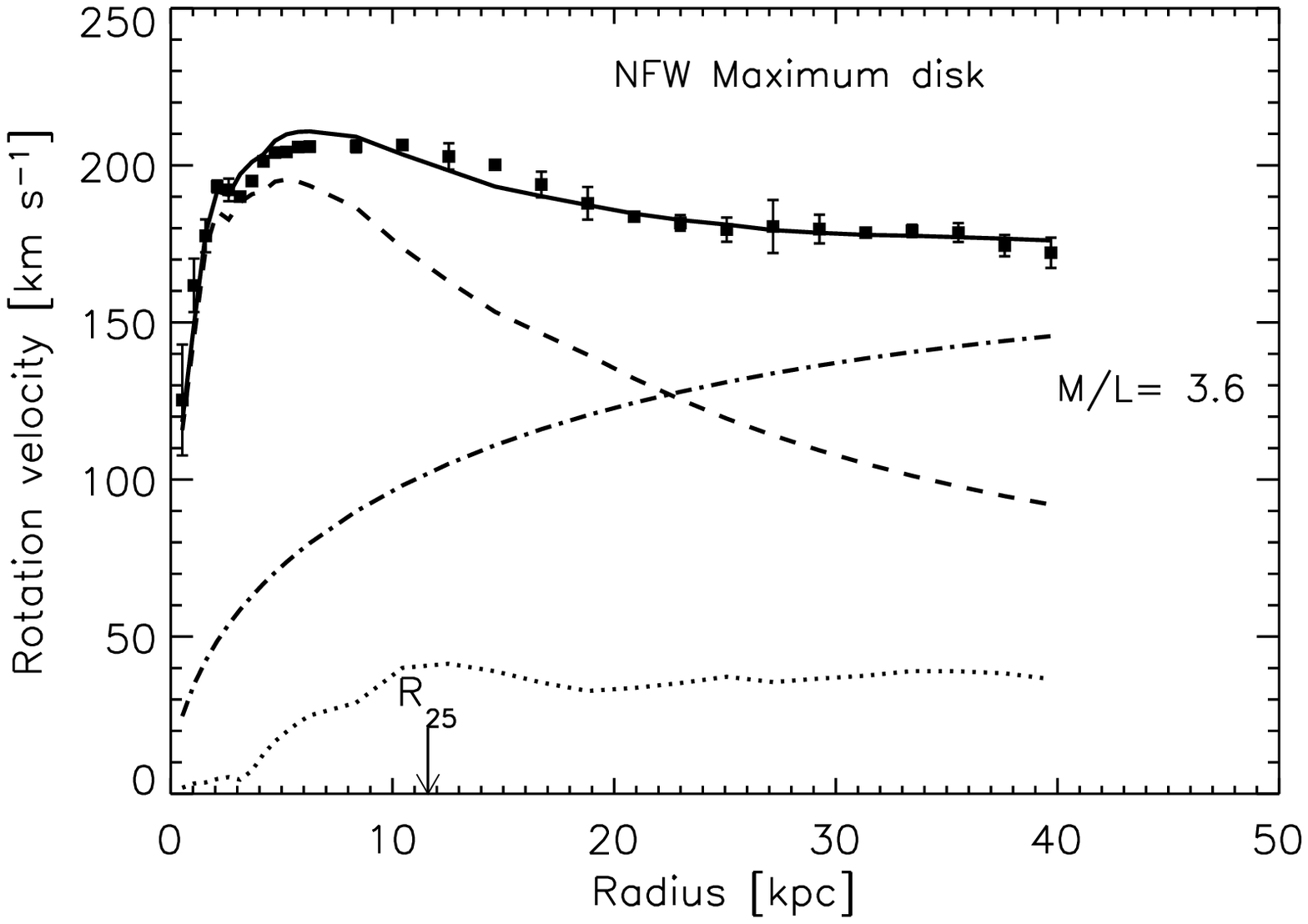}
\includegraphics[width=60mm]{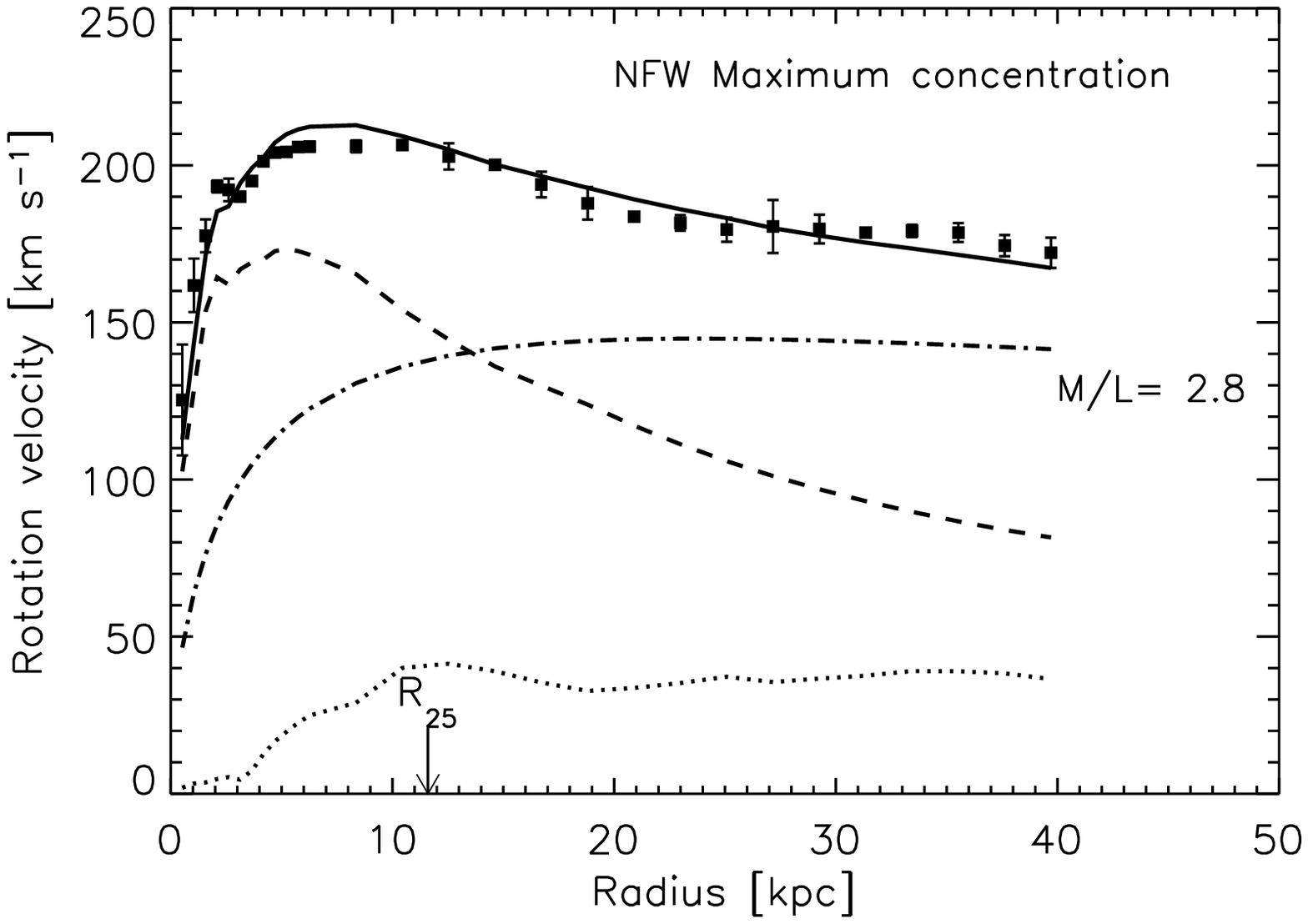}
\caption{Mass models for NGC\,5055. The contributions of the different mass components 
to the observed circular velocities are shown: gas (dotted line), stars (dashed 
line), and 
DM halo (dash-dot line). The full thick line shows the total
contribution to the rotation velocity. }
\label{fig:decomp}
\end{center}
\end{figure*}

Such a decrease in the outer parts of the rotation curve has been observed 
in other galaxies  (Casertano
and van Gorkom 1991; Bottema \& Verheijen 2002).
Interestingly, this decline happens to be just outside the 
bright optical disk ($R_{25}=$5.5$'$)  where the warp starts (Fig.~\ref{fig:parwarp_gr}) 
and the \hi\  surface density drops
abruptly (Fig.~\ref{fig:HIdensity}). 
This suggests that such a decline may be related to the ``end''
of the bright stellar disk where the dark matter halo starts to dominate. 
The decompositions of the rotation curve (Fig.~\ref{fig:decomp}) 
show that such a
decline can be reproduced by a large set of parameters: different $M/L$ 
values and different density profiles of the dark matter halo.
However for an isothermal dark halo there seems to be a lower limit for 
the $M/L$ ratio of the stellar disk. 
The top right panel of Fig.~\ref{fig:decomp} shows the fit for a 
$M/L=$1.4, which we consider as the minimum disk fit. 
Such a fit shows discrepancies between the data and the model that
are larger than in the maximum disk fit. 
For lower $M/L$ ratios than this value, the fit is unacceptable. 
The minimum disk fit obtained is very close to a so-called Bottema
disk, where 66\% of the observed maximum rotational velocity is 
contributed  by the luminous matter (Bottema 1997). 
Figure~\ref{fig:decomp} (lower panels) also shows the fits obtained 
with a universal
profile for the dark matter halo (Navarro, Frenk, \& White 1995).
 The parameter space permitted by this fit is even smaller, going from
a maximum disk ($M/L=$3.6) to a minimum disk with $M/L=$2.8. 
Note, however, that the inner rotation curve is better reproduced 
with a maximum disk fit.

In conclusion, the maximum disk fit (with either isothermal or NFW profile
for the dark component) can satisfactorily reproduce the main features
of the rotation curve of NGC\,5055 (inner bump and outer decline).
This gives a natural explanation for the decline of the rotation curve
in terms of a transition from the region where the
stellar disk is dynamically dominant and the outer parts where the dark 
halo starts to dominate. 
The existence of two such different dynamical regimes for the
inner and the outer parts of the galaxy may be related to the
outer warping of the disk (Sect.~\ref{sec:warp_disc}).

\subsection{The lopsidedness: a disk/DM halo offset?} \label{sec:symm_disc}

We have described the mild asymmetries
in both the warp and the kinematics of NGC\,5055 and 
investigated the tilted ring models and, in 
particular, the consequences of a change
in the systemic velocity and in the position of the centre from the inner 
to the outer rings of the disk. We showed that 
such a change leads to symmetrical rotation curves on the two 
halves of the disk and also to a striking symmetry for 
the position and the inclination angles on the two sides
(compare Figs.~\ref{fig:parwarp} and \ref{fig:parwarp_gr}). 
Although it may not be surprising that by allowing 
the systemic velocity to vary in each ring, symmetric rotation 
curves are obtained, it is remarkable that also the position and 
inclination angles are made symmetrical here 
(and consequently also the other angles that describe the warp).
Indeed, a simple variation with radius of centre and
systemic velocity in the tilted ring fit has led to a complete
symmetrization of the system, both kinematically and geometrically. 
It appears that we are dealing here with fundamental aspects of 
the dynamics of NGC\,5055 and that there is probably a relation with 
the potential and with the two regimes  -a stellar disk dominating 
in the inner and a dark halo in the outer parts-  discussed in the 
previous sections.
The following facts also support this view:
 
\begin{enumerate}
\item{The systemic velocity does not vary randomly.
On the contrary, it stays roughly constant in the inner region 
(within 8 kpc) and, at the end 
of the bright optical disk, rises and remains  
approximately constant around a higher value.}
\item{
The position of the centre of the galaxy obtained with the two assumed
systemic velocities also varies (by about 1.8 kpc) from
the inner to the outer rings, and towards the western side where the \hi\ 
disk of NGC\,5055 is more extended. 
The displaced outer rings resulting from this shift of the
centre roughly match the asymmetry in the \hi\ density distribution.}
\item{
The changes in systemic velocity and position of the centre 
not only produce symmetric rotation curves but also symmetrize the 
position and inclination angles and hence affect the entire 
kinematics and geometry of the galaxy.
}
\item{
The displacement and the change in systemic velocity of the outer rings 
are towards the projected position and the systemic 
velocity of the companion UGC\,8313. 
But it is unclear whether this points to a dynamical influence of the companion on the
outer parts of NGC\,5055 as already discussed in relation to the warp 
(Sect.~\ref{sec:warp_disc}).}
\end{enumerate}

In our analysis of NGC\,5055, we relaxed two of the 
basic assumptions  -the spatial 
and the kinematical centre of the system do not vary with radius-   
that are usually made in the tilted ring modeling of a galaxy.
The results are a strong indication that this is a very
powerful way to parameterize the asymmetries of the system. 
But in this approach the orbits of the gas are still circular, and the 
velocity is constant over the whole orbit. If the shifting of the centre 
with radius is related to a shifting of the mass centroid, one expects 
the orbits of the particles to be elliptical.
If the perturbation of the circular orbit is small (i.e.\ the
potential has a small stationary perturbation), an analytical
description of the motion can be given via the theory of epicycles. 

In order to test our results,
we performed a decomposition of the velocity field of NGC\,5055
into harmonic components following the approach of Schoenmakers, Franx,
and de Zeeuw \cite{schoenmakers} and found that the dominant harmonic terms 
are $m=$0 and $m=$2 (in the outer parts). 
This confirms our previous findings: the $m=$0 term can indeed be seen as a
change in systemic velocity, whilst the $m=$2 term is related to a shift
of the centre of the galaxy (see Schoenmakers et al.\ 1997).
In particular, in NGC\,5055 the $s_2$ term is significantly different
from 0 beyond 14 kpc, and this is, in fact, the radius where the
position of the centre shifts significantly from the inner value
(Fig.~\ref{fig:centre}). 

As pointed out by Schoenmakers et al. \cite{schoenmakers}, the presence of an $m$-term
stationary perturbation of the potential produces $m-$1 and $m+$1
terms in the residual velocity field.
These terms are mixed, and their reciprocal amplitudes depend mainly on
the viewing angle of the perturbation (see also Swaters et al.\ 1999). 
Our harmonic analysis suggests the presence of an $m=$1 term perturbation
of the potential in NGC\,5055.

In short, we have shown a new approach to parameterizing the
kinematical  and morphological asymmetries in the spiral galaxy
NGC\,5055 that gives the same results as the harmonic analysis
of the velocity field.
In this approach the lopsidedness is related to 
a transition from the inner parts where the dynamically dominant
component is the (axi-symmetric) stellar disk to the outer parts 
which are dominated by a displaced (or lopsided) dark matter halo.
This interpretation is supported in NGC\,5055 by the beginning of
the warp and the decline of the rotation curve at the transition
radius. 
Finally, we note that in our approach the kinematical and the 
morphological lopsidedness both 
tend to disappear as the centres of the outer rings shift in 
the direction where the \hi\ distribution is more extended. 
It is not clear whether the
presence of the companion galaxy may also play a role. 

\section{Summary}

We carried out a detailed 21-cm line study of the 
spiral galaxy NGC\,5055 using new \hi\ observations 
recently obtained with the WSRT. 
New interesting aspects have emerged concerning the morphology
and the dynamics of this galaxy: 

	1. NGC\,5055 has a very extended ($R\sim$ 40 kpc) and warped 
\hi\ disk. The warp is highly symmetric, and its long revolution 
time (1.5 Gyr) indicates that it is a long-lived
phenomenon. The warp begins at the end of the bright stellar disk and 
is oriented in the direction of the companion galaxy UGC\,8313.

	2. The rotation curve of NGC\,5055 shows two main features: a bump at
about 2 kpc from the centre, which clearly corresponds to a bump in the
optical surface brightness, and a velocity decline of about 25 \km at the
end of the bright optical disk. The standard analysis of the rotation curve 
into luminous and dark matter components shows that only "maximum disk"
fits are able to reproduce both features at the same time. Furthermore,
for an acceptable fitting of the rotation curve in its declining outer parts, 
the stellar disk must be rather massive. In the solution with a 
quasi-isothermal dark matter halo a firm lower limit has been set to the 
disk mass-to-light ratio (M/L$\gtrsim$ 1.4). 

	3. Mild asymmetries between the approaching and receding sides 
of NGC\,5055 are present both in the kinematics and in the morphology. 
The tilted ring analysis of the velocity field shows that if the systemic 
velocity and the position of the centre are allowed to vary 
with radius from the inner to the outer parts, the asymmetries are explained 
and a striking geometrical and kinematical symmetrization of the
system is achieved. 

All these results suggest that NGC\,5055 has two dynamical regimes:
an inner one dominated by the luminous matter and an outer
one, dominated by a dark matter halo offset with respect to the 
inner disk. The transition radius (about 10
kpc) is the region where:
	
\begin{itemize}
\item The warp begins
\item The stellar and the \hi\  disk begin to fade out
\item The rotation curve starts to decline
\item The systemic velocity changes
\end{itemize} 
It is not clear whether the companion galaxy UGC\,8313 has any significant 
role in causing the observed asymmetry.

\section*{Acknowledgements}
We thank Thijs van der Hulst for insightful comments 
and improvements to the manuscript. 
G.B. is grateful to the Kapteyn Astronomical Institute for their hospitality
 and financial support in the early stages of this work. 
The WSRT is operated by the Netherlands Foundation for Research in
Astronomy (ASTRON) with the support from the Netherlands Foundation for
Scientific Research (NWO).  This research made use of the NASA
Extragalactic Database (NED). The Digitized Sky Survey was produced at the
Space Telescope Science Institute under US Government grant NAG W-2166.

\bibliographystyle{aa}

\begin{thebibliography}{}

\bibitem[1987]{begeman}
Begeman, K.G. 1987, Ph.D. Thesis, University of Groningen, NL
\bibitem[2005]{boomsma}
Boomsma, R., Oosterloo, T. A., Fraternali, F., van der Hulst, J. M., Sancisi, R.
2005, A\&A, 431, 65
\bibitem[1978]{bosma}
Bosma, A. 1978, Ph.D. Thesis, University of Groningen, NL
\bibitem[1993]{bottema93}
Bottema, R. 1993, A\&A, 275, 16
\bibitem[1997]{bottema97}
Bottema, R. 1997, A\&A, 328, 517
\bibitem[2002]{bottema02}
Bottema, R., Verheijen,M.~A.~W. 2002, A\&A, 388, 793 
\bibitem[1990]{briggs}
Briggs, F.H. 1990, AJ, 352, 15
\bibitem[1992]{broeils}
Broeils, A.H. 1992, PhD Thesis, University of Groningen, NL
\bibitem[1957]{burke}
Burke, B.F. 1957, AJ, 62, 90
\bibitem[1983]{casertano83}
Casertano, S. 1983, MNRAS, 203, 735
\bibitem[1991]{casertano91}
Casertano, S. \& van Gorkom, J.H. 1991, AJ, 101, 1231
\bibitem[1980]{clark}
Clark, B.G. 1980, A\&A, 89, 377
\bibitem[1999]{debattista}
Debattista, V.P. \& Sellwood, J.A. 1999, ApJ, 513, L107
\bibitem[2002]{fraternali02}
Fraternali, F., van Moorsel, G., Sancisi, R., \& 
Oosterloo, T.  2002, AJ, 123, 3124
\bibitem[2004]{fraternali04}
Fraternali, F., Oosterloo, T., Sancisi, R., Swaters, R., 2005, in
"Extra-planar Gas", Ed. R. Braun, ASP Conference Proceedings, Vol. 331,
p.239
\bibitem[2001]{garcia}
Garcia-Ruiz, I., Sancisi R., Kuijken K., 2002, A\&A, 394, 769
\bibitem[1958]{holmberg}
Holmberg, E. 1958, Medd. Lund Obs. Ser. II no. 136
\bibitem[1969]{hunter}
Hunter, C. \& Toomre, A. 1969, ApJ, 155, 747
\bibitem[1999]{jiang}
Jiang, I. G. \& Binney, J. 1999, MNRAS, 303, L7
\bibitem[1993]{kamphuis}
Kamphuis, J. 1993, PhD Thesis, University of Groningen, NL
\bibitem[1992]{kamphuis2}
Kamphuis, J. \& Briggs, F. 1992, A\&A, 253, 335
\bibitem[1959]{kahn}
Kahn, F. D., Woltjer, L. 1959, ApJ, 130, 705
\bibitem[1987a]{kent}
Kent, S.M. 1987a, AJ, 93, 816
\bibitem[1957]{kerr}
Kerr, F.J. 1957, AJ, 62, 93
\bibitem[1996]{maoz}
Maoz, D., Filippenko, A.~V., Ho, L.~C., et al. 1996, ApJ, 107, 215
\bibitem[2003]{matthews}
Matthews, L.D., Wood, K. 2003, ApJ, 593, 721
\bibitem[1995]{navarro95}
Navarro, J.F., Frenk, C.S., \& White, S.D. 1995, MNRAS, 275, 720
\bibitem[1996]{navarro96}
Navarro, J.F., Frenk, C.S., \& White, S.D. 1996, ApJ, 462, 563
\bibitem[1991]{persic}
Persic, M., Salucci, P. 1991, ApJ, 368, 60 
\bibitem[1994]{pierce}
Pierce, M. 1994, ApJ, 430, 53
\bibitem[1980]{rots}
Rots, A.H. 1980, A\&A, 41, 189
\bibitem[2004]{sancisi}
Sancisi, R. 2004, in IAU Symp. 220, Dark Matter in Galaxies, ed. Ryder, S.D.,
Pisano, D.J.,Walker, M.A., and Freeman, K.C.,San Francisco:PASP, p.233 
\bibitem [1995]{sault}
Sault, R.~J., Teuben, P.~J. \& Wright, M.~C.~H. 1995, in Astronomical Data
Analysis Software and Systems IV, ed. R.~A. Shaw, H.~E. Payne \& J.~J.~E.
Hayes (San Francisco: ASP), ASP Conf. Ser., 77, 433
\bibitem [1997]{schoenmakers}
Schoenmakers, R.H.M., Franx, M., \& de Zeeuw, P.T. 1997, MNRAS, 292, 349
\bibitem[1985]{schwarz}
Schwarz, U.J. 1985, A\&A, 142, 273
\bibitem[1988]{sparke}
Sparke, L.S. \& Casertano, S. 1988, MNRAS, 234, 873
\bibitem[1997]{swaters97}
Swaters, R. A., Sancisi, R., \& van der Hulst, J. M.  1997, ApJ, 491, 140
\bibitem[1999]{swaters99}
Swaters, R.A., Schoenmakers, R.H.M., Sancisi, R. \& van Albada,
T.S. 1999, MNRAS, 304, 330
\bibitem[1988]{tully}
Tully, R.~B., 1988, Sci, 242, 310
\bibitem[1992]{hulst}
van der Hulst, J.~M., Terlouw, J.~P., Begeman, K., Zwitser, W., \& Roelfsema,
P. R. 1992, in Astronomical Data Analysis Software and Systems I, ed. D.~M.
Worall, C. Biemesderfer, \& J. Barnes (San Francisco: ASP), ASP Conf. Ser., 25,
131
\bibitem[1976]{vaucouleurs}
Vaucouleurs, G. de et al. 1976, Second Reference Catalogue of Bright
Galaxies, University of Texas Press, Austin
\bibitem[2002]{wechsler}
Wechsler, R.~H., Bullock, J.~S., Primack, J.~R., Kravtsov, A.V~., 
Avishai D.~A., 2002, ApJ, 568, 52
\bibitem[1984]{wevers}
Wevers, B.M.H.R. 1984, Ph.D. Thesis, University of Groningen, NL

\end{thebibliography}

\clearpage

\small
\begin{table}
\begin{center}
\begin{tabular}{lcc}
\hline
\hline
 &   &  \\
Parameter &  NGC\,5055 & Ref.   \\
  &  & \\
\hline
  &  &  \\
Morphological type & SA(rs)bc II &  1\\
Optical centre ($\alpha$, $\delta$ J2000) & 13$^h$15$^m$49.25$^s$  +42$^{\circ}$01$'$49.3$''$ & 2\\
Kinematical centre ($\alpha$, $\delta$ J2000)& 13$^h$15$^m$49.2$^s$  +42$^{\circ}$01$'$49.0$''$ & 3\\
Distance (Mpc) & 7.2 & 4\\
L$_B$ (L$_{\odot B}$) & 2.55$\times 10^{10}$ & 1\\
Disk scale length ($'$) & 1.6 & 5 \\
$R_{25}$ ($'$)  & 5.5 & 1 \\
Holmberg diametres ($'$)& 16$\times$10 & 6 \\
Systemic velocity (km s$^{-1}$)& 497.6$\pm$ 4.8 & 3 \\
$M_{\rm HI}$   & $6.2\pm0.3 \times 10^9$ \mo & 3 \\
$M_{\rm *}(<R_{\rm out})$   & $8\times 10^{10}$ \mo & 3 \\
$M_{\rm DM}(<R_{\rm out})$   & $1.9\times 10^{11}$ \mo & 3 \\
\hline
\hline
\end{tabular}
\end{center}
\caption{Optical and Radio Parameters for NGC\,5055. (1) De Vaucouleur et al., 1976; (2) Maoz et 
al., 1996; (3) this work; (4) Pierce, 1994; (5) Kent, 1987a, using a
distance of 7.2 Mpc; 
(6) Holmberg, 1958. $M_{\rm *}(<R_{\rm out})$ and $M_{\rm DM}(<R_{\rm out})$ 
are, respectively, the mass of the stellar disk and of the dark matter halo within 
the last measured point ($\simeq$ 40 kpc), 
derived in the maximum disk hypothesis with isothermal halo. 
\label{tab:data5055}}
\end{table}

\begin{table}
\begin{center}
\begin{tabular}{lc}
\hline
\hline
  &   \\
Parameter     & NGC\,5055 \\
  &  \\
\hline
 & \\
Observation dates & 22-04-2001, 11-05-2001    \\
Length of observation &  12 h each  \\
Number of antennas &      14 \\
Baseline (min-max-incr) & 36-m   2700-m   36-m \\
Pointing R.A. (J2000)& $13^h 15^m 49.20^s$ \\
Pointing DEC (J2000) & $42^{\circ} 01^{'} 48.99^{''}$ \\
Frequency of observation (MHz) & 1420.406\\
Total bandwidth (MHz) &  5 \\
Total bandwidth (km s$^{-1}$) & 948  \\
Number of channels &  128 \\
Central velocity (km s$^{-1}$) & 540 \\
Channel separation (kHz) & 38.7   \\
Channel separation (km s$^{-1}$) &  8.24\\
Spectral resolution (km s$^{-1}$) & 16.48 \\
 & \\
\hline
\hline
\end{tabular}
\caption{Observational parameters for NGC\,5055}
\label{tab:observations}
\end{center}
\end{table} 
\normalsize

\clearpage

\begin{table*}
\begin{center}
\begin{tabular}{lccc}
\hline
\hline
  &  &   &  \\  
Parameter   & 18$''$ & 28$''$ & 67$''$ \\
  &   &   &  \\
\hline
 & \\
HPBW ($''$) &   23.5$\times$13.9 & 31.6$\times$25.0 & 74.7$\times$59.3  \\
P.A. of synthesized beam ($^{\circ}$) & 1.3 & 11.2 & 10.1\\
Beam size (kpc) & 0.82$\times$0.49 & 1.10$\times$0.87 & 2.61$\times$2.07    \\
R.m.s. noise per channel (mJy/beam) & 0.3 & 0.4 & 0.6\\
R.m.s. noise per channel (K) & 0.6 & 0.3 & 0.1 \\
Minimum detectable column density:&   &   &  \\
per resolution element (cm$^{-2}$)& $8.4\times 10^{19}$ & $4.6\times 10^{19}$ & $1.2\times 10^{19}$\\
    per resolution element (M$_{\odot}/pc^2$) & 0.67 & 0.37 & 0.10\\
Conversion factor (K/mJy) & 2.0 & 0.75 & 0.17 \\
 & \\
\hline
\hline
\end{tabular}
\caption{Parameters of the Data Cubes used}
\end{center}
\label{tab:cubes}
\end{table*} 

\end{document}